\documentclass{IEEEtran}

\usepackage[cmex10]{amsmath} 
\usepackage{amssymb,amsthm}
\usepackage{mathtools,empheq}
\usepackage{graphicx,subfig,array}
\usepackage{cite}
\usepackage{enumerate}
\usepackage[usenames]{color}

\long\def\beginpgfgraphicnamed#1#2\endpgfgraphicnamed{\includegraphics{#1}}

\newtheorem{theorem}{Theorem}
\newtheorem{lemma}{Lemma}

\usepackage{varioref}
\usepackage[breaklinks=true,%
            colorlinks=true,
            linkcolor=black,
            anchorcolor=black,
            citecolor=black,%
            filecolor=black,%
            menucolor=black,%
            pagecolor=black,%
            urlcolor=black,%
            plainpages=false,%
            bookmarksopen=true,
            bookmarksnumbered=true 
           ]{hyperref}

\newcommand{\OO}{\ensuremath{\mathrm{O}}}

\newcommand{\FF}{\ensuremath{\mathbb{F}}}

\newcommand{\VV}{\ensuremath{V}}
\newcommand{\VVint}{\ensuremath{\overset{\circ}{V}}}

\newcommand{\NN}{\ensuremath{N}}

\newcommand{\uu}{\ensuremath{u}}
\newcommand{\vv}{\ensuremath{v}}
\newcommand{\ww}{\ensuremath{w}}

\newcommand{\vvm}[1]{\ensuremath{v^{\setminus{#1}}	}}
\newcommand{\uum}[1]{\ensuremath{u^{\setminus{#1}}	}}

\newcommand{\ie}{{\it i.e.},\ }

\newcommand{\dist}{\ensuremath{\mathrm{dist}}}

\newcommand{\Enc}{\ensuremath{\mathcal{E}_{\mathrm{coding}}}}
\newcommand{\Er}{\ensuremath{\mathcal{E}_{\mathrm{routing}}}}

\newcommand{\Bfix}{\ensuremath{B_{\mathrm{fixed}}}}
\newcommand{\Bvar}{\ensuremath{B_{\mathrm{var}}}}

\begin{document}
\suppressfloats[t]

\title{Lower Bounds on the Maximum Energy Benefit of\\ Network Coding for Wireless Multiple Unicast
}

\author{
Jasper Goseling, Ryutaroh Matsumoto, Tomohiko Uyematsu and Jos H.\ Weber\thanks{J. Goseling and J.H. Weber are with Delft University of Technology, Mekelweg 4, 2628 CD Delft, The Netherlands, email: \url{j.goseling@tudelft.nl}, \url{j.h.weber@tudelft.nl}. R. Matsumoto and T. Uyematsu are with Tokyo Institute of Technology, Tokyo, 152--8550 Japan, email: \url{ryutaroh@rmatsumoto.org}, \url{uyematsu@ieee.org}. J. Goseling is also with University of Twente, The Netherlands. Parts of this work have been presented at the 2008 Symposium on Information Theory in the Benelux and the 2009 IEEE International Symposium on Information Theory.}
}

\maketitle

%
%
%
\begin{abstract}
\noindent We consider the energy savings that can be obtained by employing network coding instead of plain routing in wireless multiple unicast problems. We establish lower bounds on the benefit of network coding, defined as the maximum of the ratio of the minimum energy required by routing and network coding solutions, where the maximum is over all configurations. It is shown that if coding and routing solutions are using the same transmission range, the benefit in $d$-dimensional networks is at least $2d/\lfloor\sqrt{d}\rfloor$. Moreover, it is shown that if the transmission range can be optimized for routing and coding individually, the benefit in $2$-dimensional networks is at least $3$. Our results imply that codes following a \emph{decode-and-recombine} strategy are not always optimal regarding energy efficiency.
\end{abstract}

%
%
%
\section{Introduction} \label{sec:intro}
Emerging applications in wireless networks, like environment monitoring in rural areas by adhoc networks, require more and more resources. One of the most important limitations is formed by battery life. Since battery technology is not keeping up with the increasing demand from resource-consuming applications, it is imperative that more efficient use is made of the available energy. There has been significant recent attention to the problem of minimizing energy consumption in networks. Some of the topics considered are minimum cost routing~\cite{chen98qosoverview, chang98energy, rodoplu99}, power control algorithms~\cite{ramanathan02topology, jones01survey, saraydar02efficient} and cross-layer protocol design for energy minimization~\cite{goldsmith02challenges}. In this work we are interested in the use of network coding~\cite{ahlswede00,li03linear,koetter03algebraic,ho06,nctheory,ncmonol1,ncmonol2} for reducing the energy consumption in wireless networks. We compare the reduction with traditional routing solutions. The contributions of this work are lower bounds on the energy reduction that can be achieved by using network coding for multiple unicast problems in wireless networks.

In recent years there has been significant interest in network coding with the aim of reducing energy consumption in networks. More generally, network coding with a cost criterion has been considered. Much progress has been made in understanding the case of multicast traffic. In fact, it has been shown by Lun et al.\ that a minimum-cost network coding solution can be found in a distributed fashion in polynomial time~\cite{lun06mincost}. The fact that the complexity of finding this solution is polynomial in time is surprising, since the corresponding routing problem is a Steiner tree problem that is known to be NP-complete~\cite{winter87steiner}.

Besides constructing minimum-cost coding solutions it is also of interest to know what the benefits of network coding are compared to routing. In this work we are interested in the energy benefit of network coding, which is the ratio of the minimum energy solution in a routing solution compared to the minimum energy network coding solution, maximized over all configurations. It has been shown by Goel and Khanna~\cite{goel08bound} that the energy benefit of network coding for multicast problems in wireless networks is upper bounded by a constant. The problem of reducing energy consumption for many-to-many broadcast traffic in wireless networks has been studied by Fragouli et al.\ in~\cite{fragouli08ebu} and Widmer and Le Boudec in~\cite{widmer05extreme}, providing lower bounds on the energy benefit of network coding for specific topologies. More importantly, algorithms have been presented in~\cite{fragouli08ebu,widmer05extreme} that allow to exploit these benefits in practical scenarios, \ie in a distributed fashion.

The above demonstrates that for multicast traffic and for many-to-many broadcast traffic, there is some understanding of the energy benefits of network coding and how to exploit them. In order to reduce energy consumption in practical networks, however, it is important to consider also multiple unicast traffic. Indeed, in practice a large part of the data will be generated by unicast sessions. For the case of multiple unicast traffic, contrary to multicast and broadcast, not much is known. This paper deals with the energy benefits of network coding for wireless multiple unicast. Remember from the above that for multicast, the problem of minimum-cost routing is hard, whereas minimum-cost network coding is easy. In stark contrast, the problem of minimum-cost multiple unicast routing is easy. One constructs the minimum-cost solution, \ie the shortest path, for each session individually. The minimum-cost multiple unicast network coding problem, however, seems hard and in general very little is known. 

Network coding for the multiple unicast problem was first studied by Wu et al. in~\cite{wu05ciss}, in which it was shown that in the information exchange problem on the line network the energy saving achieved by network coding is a factor two. The network codes that we construct in this work are in a sense a generalization of the results on one dimensional networks~\cite{wu05ciss}, to higher dimensional networks. The networks considered in this work are lattices. More specifically, the hexagonal lattice and the rectangular lattice. Effros et al.~\cite{effros06tiling} and Kim et al.\cite{kim07low} have considered energy-efficient network codes on the hexagonal lattice. We improve the lower bounds on the energy savings of network on the hexagonal lattice given in~\cite{effros06tiling}. More precisely, we improve the previously known bound of $2.4$ and obtain a new bound of $3$.

Kramer and Savari have developed techniques that can be used to upper bound the achievable throughputs in a multiple unicast problem~\cite{kramer06ecb}. No methods are known, however, to lower bound the cost of network coding solutions for a configuration. A lower bound to the ratio of the minimum energy consumption of routing and coding solutions for a given multiple unicast configuration was provided by Keshavarz-Haddad and Riedi in~\cite{keshavarzhaddad08}. For the type of configurations used in this paper, however, the results from~\cite{keshavarzhaddad08} give the trivial lower bound of one. We will see, however, that network coding has large energy savings for these configurations.

An important class of network codes operates according to a principle that we will refer to in the remainder as \emph{decode-and-recombine}. These codes satisfy the constraint that each symbol in each linear combination that is transmitted is explicitly known by the node transmitting that linear combination. Note, that this is a restriction from the general linear coding strategy, in which linear combinations of coded messages can be retransmitted. The motivation behind using decode-and-recombine codes is that it prevents information from spreading too much in the network, away from the path between source and destination, a heuristic introduced by Katti et al.~\cite{katti06xor}. The use of a decode-and-recombine strategy results in reduced complexity. However, an important question that has to be addressed is, whether the use of decode-and-recombine codes leads to a higher energy consumption than is strictly necessary. We answer this question affirmatively. An upper bound of three on the energy benefit of decode-and-recombine codes has been given by Liu et al.~\cite{liu07scale}. One of the contributions of this work is to show that larger energy benefits can be obtained by considering also other types of codes.

This paper is organized as follows. In Section~\ref{sec:model} we specify our model and problem statement more precisely. Our main results are presented in Section~\ref{sec:results}. Constructions of configurations that allow a large energy benefit for network coding and proofs of our results are given in Sections~\ref{sec:hex} and~\ref{sec:rect}. In Section~\ref{sec:concl}, finally, we discuss our work.

%
%
%
\section{Model and Problem Statement} \label{sec:model}
Let $\VV\subset\mathbb{R}^d$ be the nodes of a $d$-dimensional wireless network.
We consider a wireless network model with broadcast, where all nodes within range $r$ of a transmitting node can receive, and nodes outside this range cannot. More precisely, given a transmission range $r$, a node $\vv$ is broadcasting to all nodes in the set 
\begin{equation*}
\{\uu\in\VV\ |\ \|\uu-\vv\|\leq r\},
\end{equation*}
where $\|\uu-\vv\|$ denotes the Euclidean norm of $\uu-\vv$. The energy required to transmit one unit of information to all other nodes within range $r$ equals $cr^\alpha$, where $\alpha$ is the path loss exponent and $c$ some constant. In analyzing the energy consumption of nodes, we will consider only the energy consumed by transmitting. Receiver energy consumption as well as energy consumed by processing are assumed to be negligible compared to transmitter energy consumption. In particular, note that little additional processing is required for network coding, compared to the processing that is performed in a traditional wireless protocol stack.

The traffic pattern that we consider is multiple unicast. All symbols are from the field $\FF_2$, \ie they are bits and addition corresponds to the xor operation. The source of each unicast session has a sequence of source symbols that need to be delivered to the corresponding destination. Let $M$ be the set of unicast sessions. We call $\{\VV,M,r\}$ a wireless multiple unicast configuration.

We will compare energy consumption of routing and network coding. Our goal is to establish lower bounds on the maximum of the ratio of the minimum energy required by routing and network coding solutions, where the maximum is over all configurations. We will refer to this ratio as the \emph{energy benefit of network coding}. Let $\Enc(V,M,r)$ and $\Er(V,M,r)$ be the minimum energy required for network coding and routing solutions, respectively, for a configuration $\{V,M,r\}$. The energy consumption of a coding or routing scheme is defined as the time-average of the total energy spent by all nodes in the network to deliver one symbol for each unicast session. In analyzing coding schemes we will ignore the energy consumption in an initial startup phase and consider only steady-state behavior.

Note that since energy consumption per transmission equals $cr^\alpha$, the transmission range $r$ is an important factor in the energy consumption. Therefore, it is of particular interest to optimize the transmission range such that energy consumption is minimized. In this work we consider two different quantities: 1) $\Bfix$, denoting the energy benefit that can be obtained if the transmission range is given and fixed and 2) $\Bvar$, denoting the energy benefit that can be obtained if one is allowed to optimize the transmission range. Note that the transmission range can be individually optimized for the routing and network coding scenarios. More precisely, the goal of this work is to establish lower bounds on 
\begin{equation*}
\Bfix(d)
 = \max_{V,M,r}
\frac{\Er(V,M,r)}{\Enc(V,M,r)},
\end{equation*}
where the maximization is over all node locations $V\subset\mathbb{R}^d$, multiple unicast sessions $M$ and transmission ranges $r$, with the transmission range equal for the routing and network coding solutions, and
\begin{equation*}
\Bvar(d)
 = \max_{V,M}
\frac{\min_r \Er(V,M,r)}{\min_r \Enc(V,M,r)},
\end{equation*}
where the maximization is over all node locations $V\subset\mathbb{R}^d$ and multiple unicast sessions $M$, with the transmission range optimized individually for the routing and network coding solutions. If no confusion can arise, we will omit dependency on $d$ in the notation for $\Bfix$ and $\Bvar$.

Since in $\Bfix$, $r$ is equal for $\Er$ and $\Enc$, the energy per transmission is equal in $\Er$ and $\Enc$ and the benefit is equal to the ratio of the number of transmissions required in routing and network coding solutions.

Since we are interested in energy consumption only, we can assume that all transmissions are scheduled sequentially and/or that there is no interference. All coding and routing schemes that we consider proceed in time slots or rounds. In each time slot all nodes are allowed to transmit one or more messages. We assume that the length of the time slot is large enough to accommodate sequential transmission of all messages in that round. Coding operations will be based on messages received in previous time slots only. Finally, we assume that all nodes have complete knowledge of the network topology and the network code that is being used.

To conclude this section, we introduce here some of the notation that will be used in the remainder of the paper. The symbol transmitted by a node $\vv\in V$ in time slot $t$ is denoted by $x_t(\vv)$. If $\vv$ transmits more than one symbol in time slot $t$, these will be distinguished by a superscript, giving, for instance, $x^1_t(\vv)$ and $x^2_t(\vv)$. Nodes are represented by vectors. Given vectors $\uu=(u_1,\dots,u_d)$ and $\vv=(v_1,\dots,v_d)$, let $u_k^l\triangleq(u_k,\dots,u_l)$, $(\uu,\vv)\triangleq(u_1,\dots,u_d,v_1,\dots,v_d)$ and $\uum{i}\triangleq(u_1,\dots,u_{i-1},u_{i+1},\dots,u_d)=(u_1^{i-1},u_{i+1}^d)$.

Unicast sessions are denoted by $m^i(\uu)$, with $i$ an integer and $\uu$ a vector. We will see in Sections~\ref{sec:hex} and~\ref{sec:rect} that $\uu$ defines the location of the source and $i$ the relative location of the destination, \ie the direction of the session. In some cases $m^i(\uu)$ will be denoted as $m^i(u_1,u_2^d)$ or similar forms. The $t$-th source symbol of a session $m^i(\uu)$ is denoted by $m_t^i(\uu)$. The source and destination of session $m^i(\uu)$ are denoted by $s^i(\uu)$ and $r^i(\uu)$ respectively.

%
%
%
\section{Results} \label{sec:results}
We provide lower bounds on $\Bvar$ and $\Bfix$.
\begin{theorem} \label{th:fix}
The ratio of the minimum energy consumption of routing solutions and the minimum energy consumption of network coding solutions, maximized over all node locations, multiple unicast sessions and transmission ranges, with the transmission range equal for the routing and network coding solutions, is at least $2d/\lfloor\sqrt{d}\rfloor$, \ie
\begin{equation*}
\Bfix(d) \geq \frac{2d}{\left\lfloor\sqrt{d}\right\rfloor}.
\end{equation*}
\end{theorem}
The result states that $\Bfix$ is at least $2$, $4$ and $6$ for $1$, $2$ and $3$-dimensional networks, respectively. The result that $\Bfix$ is at least $2$ in one dimensional networks also follows from the results in~\cite{wu05ciss}. The lower bound $4$ for $2$-dimensional networks exceeds the previously known bound of $2.4$~\cite{effros06tiling}. This new lower bound is of particular interest, since it exceeds the upper bound of $3$ for decode-and-recombine type network codes~\cite{liu07scale}. Indeed, the code that we construct does not follow a decode-and-recombine strategy. This shows that energy can be saved by considering strategies other than decode-and-recombine. No lower bounds for three dimensional networks have been previously established.

Before proving Theorem~\ref{th:fix} in Section~\ref{sec:rect} we provide some intuition. The configuration used to proof Theorem~\ref{th:fix} has nodes placed at a $d$-dimensional rectangular lattice, connectivity $r=\sqrt{d}$ and is parametrized by an integer $K$ controlling the size of the network. The network is given in Figure~\ref{fig:rectintro} for $d=2$ and $K=5$. For $d=2$ the result of Theorem~\ref{th:fix} is obtained as follows.  First consider the case of routing. Note, that the minimum-energy solution is to route all packets along the shortest path between source and destination. Therefore, all nodes in the interior of the network will need to transmit four times. Now, for the case of network coding, we will show in Section~\ref{sec:rect} that it is possible to construct a network code in which each node in the interior of the network is transmitting only once in each time slot. Therefore, by considering large $K$ and neglecting the energy consumption at the borders of the network the obtained energy benefit is $4$. 

\begin{figure}
\centering
\subfloat[]{
\beginpgfgraphicnamed{pgffigrectlatticeintro}
\begin{tikzpicture}[scale=.6]
\tikzstyle{vertex}=[circle,fill,minimum size=5pt, inner sep=0pt]

\setcounter{tikzA}{5}

\foreach \x in {0,...,\thetikzA}
	\foreach \y in {0,...,\thetikzA}
		\node[vertex] (x\x y\y) at (\x,\y) {};

\setcounter{tikzB}{\thetikzA-1}
\foreach \x in {0,...,\thetikzB}
	{
	\setcounter{tikzC}{\x+1}
	\draw (x\x y0) to (x\thetikzC y0);
	\draw (x\x y0) to (x\thetikzC y1);
	\draw (x\x y0) to (x\x y1);
	\foreach \y in {1,...,\thetikzB}
		{
		\setcounter{tikzD}{\y-1}
		\setcounter{tikzE}{\y+1}
		\draw (x\x y\y) to (x\thetikzC y\thetikzD);
		\draw (x\x y\y) to (x\thetikzC y\y);
		\draw (x\x y\y) to (x\thetikzC y\thetikzE);
		\draw (x\x y\y) to (x\x y\thetikzE);
		}
	\draw (x\x y\thetikzA) to (x\thetikzC y\thetikzA);
	\draw (x\x y\thetikzA) to (x\thetikzC y\thetikzB);
	\draw (x\thetikzA y\x) to (x\thetikzA y\thetikzC);
	}

\draw[decorate,decoration=brace] (0,5.5) -- node[above] {\scriptsize\mbox{$K+1$ nodes}} (5,5.5);
\path (-0.5,0) -- (5.5,0); 
\end{tikzpicture}
\endpgfgraphicnamed
\label{fig:rectlatticeintro}
} 
\subfloat[]{
\beginpgfgraphicnamed{pgffigrectsessionsintro}
\begin{tikzpicture}[scale=.6]
\tikzstyle{vertex}=[circle,fill, minimum size=5pt, inner sep=0pt]

\setcounter{tikzA}{5}

\foreach \x in {0,...,\thetikzA}
	\foreach \y in {0,...,\thetikzA}
		\node[vertex] (x\x y\y) at (\x,\y) {};

\setcounter{tikzB}{\thetikzA-1}
\foreach \y in {1,...,\thetikzB}
	{
	\draw[<->] (x0y\y) to (x\thetikzA y\y);
	\draw[<->] (x\y y0) to (x\y y\thetikzA);
	}

\path (-0.5,0) -- (5.5,0); 
\end{tikzpicture}
\endpgfgraphicnamed
\label{fig:rectsessionsintro}
} 
\caption{Configuration for which $\Er/\Enc=2d/\lfloor\sqrt{d}\rfloor$, with $d=2$ depicted here, is achievable. Nodes are located at integer coordinates in a $d$-dimensional space, with connectivity given by $r=\sqrt{d}$, as depicted in~\subref{fig:rectlatticeintro}. Unicast sessions are placed according to~\subref{fig:rectsessionsintro}.\label{fig:rectintro}}

\centering
\subfloat[]{
\beginpgfgraphicnamed{pgffighexlatticeintro}
\begin{tikzpicture}[scale=.6,x={(1cm,0cm)},y={(0.5cm,0.87cm)}]
\tikzstyle{vertex}=[circle,fill,minimum size=5pt, inner sep=0pt]

\setcounter{tikzA}{5}

\foreach \y in {0,...,\thetikzA}
	\setcounter{tikzB}{\thetikzA-\y}
	\foreach \x in {0,...,\thetikzB}
		\node[vertex] (x\x y\y) at (\x,\y) {};

\setcounter{tikzC}{\thetikzA-1}
\foreach \y in {0,...,\thetikzC}
	\setcounter{tikzB}{\thetikzC-\y}
	\foreach \x in {0,...,\thetikzB}
		\setcounter{tikzD}{\y+1}
		\draw (x\x y\y) to (x\x y\thetikzD);

\setcounter{tikzD}{\thetikzA-1}
\foreach \y in {0,...,\thetikzD}
	\setcounter{tikzB}{\thetikzD-\y}
	\foreach \x in {0,...,\thetikzB}
		\setcounter{tikzC}{\x+1}
		\draw (x\x y\y) to (x\thetikzC y\y);

\setcounter{tikzD}{\thetikzA-1}
\foreach \y in {0,...,\thetikzD}
	\setcounter{tikzB}{\thetikzD-\y}
	\foreach \x in {0,...,\thetikzB}
		\setcounter{tikzC}{\x+1}
		\setcounter{tikzE}{\y+1}
		\draw (x\x y\thetikzE) to (x\thetikzC y\y);

\path (-0.5,0) -- (5.5,0); 
\end{tikzpicture}
\endpgfgraphicnamed
\label{fig:hexlatticeintro}
} 
\subfloat[]{
\beginpgfgraphicnamed{pgffighexsessionsintro}
\begin{tikzpicture}[scale=.6,x={(1cm,0cm)},y={(0.5cm,0.87cm)}]
\tikzstyle{vertex}=[circle,fill, minimum size=5pt, inner sep=0pt]

\setcounter{tikzA}{5}

\foreach \y in {0,...,\thetikzA}
	\setcounter{tikzB}{\thetikzA-\y}
	\foreach \x in {0,...,\thetikzB}
		\node[vertex] (x\x y\y) at (\x,\y) {};

\setcounter{tikzB}{\thetikzA-1}
\foreach \y in {1,...,\thetikzB}
	{
	\setcounter{tikzC}{\thetikzA-\y}
	\draw[->] (x0y\y) to (x\thetikzC y\y);
	}

\setcounter{tikzB}{\thetikzA-1}
\foreach \x in {1,...,\thetikzB}
	{
	\setcounter{tikzC}{\thetikzA-\x}
	\draw[->] (x\x y\thetikzC) to (x\x y0);
	}

\setcounter{tikzB}{\thetikzA-1}
\foreach \x in {1,...,\thetikzB}
	\draw[->] (x\x y0) to (x0y\x);

\path (-0.5,0) -- (5.5,0); 
\end{tikzpicture}
\endpgfgraphicnamed
\label{fig:hexsessionsintro}
} 
\caption{Configuration for which $\Er/\Enc=3$ is achievable. Nodes are a subset of the hexagonal lattice, with connectivity as depicted in~\subref{fig:hexlatticeintro}. Unicast sessions are placed according to~\subref{fig:hexsessionsintro}.\label{fig:hexintro}}
\end{figure}
In Section~\ref{sec:rect} we will consider the general case of arbitrary $d$. Again, the network coding solution will be such that each of the $K^d+\OO(K^{d-1})$ nodes in the interior of the network is transmitting only once in each time slot. In analyzing the routing solution some care needs to the taken. Since $r=\sqrt{d}$, the number of hops that need to be taken on the shortest path between source and destination equals $\lceil K/\lfloor\sqrt{d}\rfloor\rceil$. By noting that the number of sessions is roughly equal to the number of nodes at the border of the network, \ie $2dK^{d-1}+\OO(K^{d-2})$, and ignoring all transmission from nodes at the border of the network, we establish
\begin{align}
\Bfix(d)
&\geq \lim_{K\to\infty}\frac{\left[(2dK^{d-1}+\OO(K^{d-2})\right]\left\lceil K/\lfloor\sqrt{d}\rfloor\right\rceil}{K^d+\OO(K^{d-1})} \\
&= \lim_{K\to\infty}\frac{2d/\lfloor\sqrt{d}\rfloor K^d + \OO(K^{d-1})}{K^d + \OO(K^{d-1})} \\
&= \frac{2d}{\left\lfloor\sqrt{d}\right\rfloor}.
\end{align}
Details of the configuration and a proof of Theorem~\ref{th:fix} are given in Section~\ref{sec:rect}.

The configuration and network code construction used for Theorem~\ref{th:fix} are not useful for obtaining bounds on $\Bvar$. Since, $r=\sqrt{d}$, the cost per transmission in the network coding scheme is $cd^{\alpha/2}$. One can verify, however, that the optimal transmission range under routing is $r=1$. This requires $K$ hops per session, with the cost per transmission equal $c$. Using the network code described above and the optimal routing solution at $r=1$ gives
\begin{align}
\Bvar(d)
&\geq \lim_{K\to\infty}\frac{cK\left[(2dK^{d-1}+\OO(K^{d-2})\right]}{cd^{\alpha/2}\left[K^d+\OO(K^{d-1})\right]} \\
&= 2d^{1-\alpha/2},
\end{align}
which is at most $2$, since $\alpha\geq 2$. Note that it was already shown in~\cite{wu05ciss} that $\Bvar(1)\geq 2$ and in~\cite{effros06tiling} that $\Bvar(2)\geq 2.4$.

By considering a different configuration we show that $\Bvar(2)\geq 3$.
\begin{theorem} \label{th:var}
For $2$-dimensional wireless networks, the ratio of the minimum energy consumption of routing solutions and the minimum energy consumption of network coding solutions, maximized over all node locations and multiple unicast sessions, with the transmission range optimized individually for the routing and network coding solutions, is at least $3$, \ie
$\Bvar(2)\geq 3.$
\end{theorem}
Here we provide an intuitive explanation of this result; details of the configuration and a proof of Theorem~\ref{th:var} are provided in Section~\ref{sec:hex}. The result is established using a multiple unicast configuration on a subset of the $2$-dimensional hexagonal lattice as depicted in Figure~\ref{fig:hexintro}. The minimum cost routing solution on this network follows shortest paths for all sessions and will require all nodes in the interior of the network to transmit three times in order to deliver one symbol for each session. In Section~\ref{sec:hex} we construct a network code in which each node in the interior is only transmitting once per delivered symbol. By making the size of the network large, the influence of the borders becomes negligible. Hence, the energy benefit is $3$.

Besides providing new lower bounds on the energy benefit of network, the network codes that are constructed in this paper are of interest by themselves. They might lead to insight in how to operate in networks with another structure. Finally, even though the case $d>3$ is not of any practical relevance, the bounds as well as the code constructions might lead to a better insight for lower dimensional networks.

%
%
%
\section{An Efficient Code on the Hexagonal Lattice} \label{sec:hex}
\begin{figure}
\begin{center}
\beginpgfgraphicnamed{pgffighexlattice}
\begin{tikzpicture}[scale=1,x={(1cm,0cm)},y={(0.5cm,0.87cm)}]
\tikzstyle{vertex}=[circle,fill, minimum size=8pt, inner sep=0pt]

\setcounter{tikzA}{5}

\foreach \y in {0,...,\thetikzA}
	\setcounter{tikzB}{\thetikzA-\y}
	\foreach \x in {0,...,\thetikzB}
		\node[vertex] (x\x y\y) at (\x,\y) {};

\setcounter{tikzC}{\thetikzA-1}
\foreach \y in {0,...,\thetikzC}
	\setcounter{tikzB}{\thetikzC-\y}
	\foreach \x in {0,...,\thetikzB}
		\setcounter{tikzD}{\y+1}
		\draw (x\x y\y) to (x\x y\thetikzD);

\setcounter{tikzD}{\thetikzA-1}
\foreach \y in {0,...,\thetikzD}
	\setcounter{tikzB}{\thetikzD-\y}
	\foreach \x in {0,...,\thetikzB}
		\setcounter{tikzC}{\x+1}
		\draw (x\x y\y) to (x\thetikzC y\y);

\setcounter{tikzD}{\thetikzA-1}
\foreach \y in {0,...,\thetikzD}
	\setcounter{tikzB}{\thetikzD-\y}
	\foreach \x in {0,...,\thetikzB}
		\setcounter{tikzC}{\x+1}
		\setcounter{tikzE}{\y+1}
		\draw (x\x y\thetikzE) to (x\thetikzC y\y);

\node[anchor=east] at (x0y0.west) {\scriptsize\mbox{$(0,0)$}};
\node[anchor=north] at (x1y0.south) {\scriptsize\mbox{$(1,0)$}};
\node[anchor=east] at (x0y1.west) {\scriptsize\mbox{$(0,1)$}};
\node[anchor=north] at (x\thetikzA y0.south) {\scriptsize\mbox{$(K,0)$}};
\node[anchor=east] at (x0y\thetikzA .west) {\scriptsize\mbox{$(0,K)$}};
\end{tikzpicture}
\endpgfgraphicnamed
\caption{Nodes at a subset of the hexagonal lattice with the connectivity induced by a transmission range $r=1$. The size of the network is controlled by $K$, with $K=5$ in this figure. \label{fig:hexlattice}}
\end{center}
\end{figure}
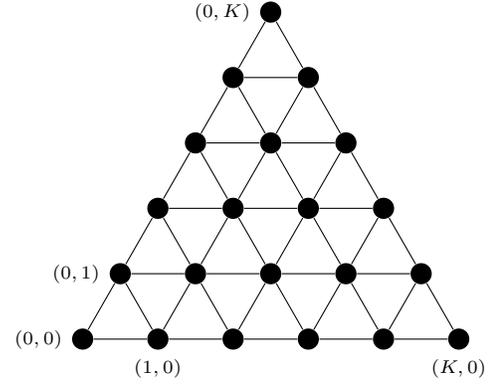
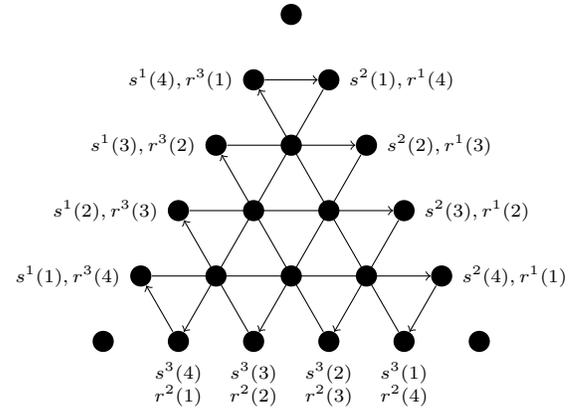
\begin{figure}
\begin{center}
\beginpgfgraphicnamed{pgffighexsessions}
\begin{tikzpicture}[scale=1,x={(1cm,0cm)},y={(0.5cm,0.87cm)}]
\tikzstyle{vertex}=[circle,fill, minimum size=8pt, inner sep=0pt]

\setcounter{tikzA}{5}

\foreach \y in {0,...,\thetikzA}
	\setcounter{tikzB}{\thetikzA-\y}
	\foreach \x in {0,...,\thetikzB}
		\node[vertex] (x\x y\y) at (\x,\y) {};

\setcounter{tikzB}{\thetikzA-1}
\foreach \y in {1,...,\thetikzB}
	{
	\setcounter{tikzC}{\thetikzA-\y}
	\draw[->] (x0y\y) to (x\thetikzC y\y);
	\node[anchor=east] at (x0y\y.west) {\scriptsize\mbox{$s^1(\y), r^3(\thetikzC)$}};
	}

\setcounter{tikzB}{\thetikzA-1}
\foreach \x in {1,...,\thetikzB}
	{
	\setcounter{tikzC}{\thetikzA-\x}
	\draw[->] (x\x y\thetikzC) to (x\x y0);
	\node[anchor=west] at (x\x y\thetikzC.east) {\scriptsize\mbox{$s^2(\x), r^1(\thetikzC)$}};
	\node[anchor=north] at (x\x y0.south) {\scriptsize\mbox{$\begin{array}{c}s^3(\thetikzC)\\ r^2(\x)\end{array}$}};
	}

\setcounter{tikzB}{\thetikzA-1}
\foreach \x in {1,...,\thetikzB}
	\draw[->] (x\x y0) to (x0y\x);

\end{tikzpicture}
\endpgfgraphicnamed
\caption{The multiple unicast sessions on the network from Figure~\ref{fig:hexlattice}. \label{fig:hexsessions}}
\end{center}
\end{figure}

In this section we present a multiple unicast configuration in which the nodes form a subset of the hexagonal lattice. It will be shown that the energy benefit on this configuration is $3$, proving Theorem~\ref{th:var}. Since the code construction used here is less involved then the construction used to prove Theorem~\ref{th:fix}, we start with the proof of Theorem~\ref{th:var}. This section is organized as follows. In Subsection~\ref{sec:hexconf} we present the configuration in more detail after which we give the construction of the network code in Subsection~\ref{sec:hexconstr}. Subsection~\ref{sec:hexvalid} is used to prove that the code is valid. Finally, in Subsection~\ref{sec:hexconcl} we analyze the energy consumption of the network code and prove Theorem~\ref{th:var}.

\subsection{Configuration} \label{sec:hexconf}
The size of the configuration is parametrized by a positive integer $K$. The nodes $V$ form a subset of the hexagonal lattice. We index nodes with a tuple $(v_1,v_2)\in\mathbb{N}^2$. $V$ is given by
\begin{equation}
\VV = \big\{(v_1,v_2) \big| v_1,v_2\geq 0, v_1,v_2\leq K, v_1+v_2\leq K \big\}.
\end{equation}
The location of node $v\in\VV$ in $\mathbb{R}^2$ is given by $v G$, where
\begin{equation}
G=\begin{bmatrix} 1 & 0 \\ 1/2 & \sqrt{3}/2  \end{bmatrix}.
\end{equation}
Let $\VVint$ denote the interior of the network, \ie
\begin{equation}
\VVint = \left\{v\in \VV \mid v_1,v_2> 0, v_1,v_2<K, v_1+v_2<K \right\}.
\end{equation}
The transmission range that we are interested in is $r=1$. This leads to connectivity between the six nearest neighbours. Hence, the neighbours of a node $(u_1,u_2)\in\VVint$ are
\begin{multline*}
	(u_1-1,u_2+1), (u_1,u_2+1), (u_1-1,u_2),\\ (u_1+1,u_2), (u_1,u_2-1), (u_1+1,u_2-1).
\end{multline*}
The nodes $V$ and the connectivity are depicted in Figure~\ref{fig:hexlattice}.

There are $3(K-1)$ unicast sessions, denoted by $m^1(i)$, $m^2(i)$ and $m^3(i)$, $1\leq i\leq K-1$. Sources and destinations of the sessions are positioned as follows
\begin{alignat}{2}
	m^1(i):\quad s^1(i) &= (0,i),   & r^1(i) &= (K-i,i) \\
	m^2(i):\quad s^2(i) &= (i,K-i), & r^2(i) &= (i,0) \\
	m^3(i):\quad s^3(i) &= (K-i,0), & r^3(i) &= (0,K-i),
\end{alignat}
as depicted in Figure~\ref{fig:hexsessions}. Remember from Section~\ref{sec:model}, that session $m^j(i)$ has the sequence of source symbols $m^j_0(i), m^j_1(i), m^j_2(i), \dots$ to be transferred.

\begin{figure*}
\centering
\subfloat[$t=0$]{
\beginpgfgraphicnamed{pgffighext0}
\begin{tikzpicture}[scale=1,x={(15mm,0cm)},y={(7.5mm,15mm)}]
\tikzstyle{vertex}=[draw, rounded corners=2pt, minimum width=13mm, minimum height=13mm]
\tikzstyle{use}=[very thick,loosely dashed]
\setlength{\extrarowheight}{2.5pt}
\setcounter{tikzA}{4}

\node[vertex] (x0y0) at (0,0) {};
\node[vertex] (x1y0) at (1,0) {};
\node[vertex] (x2y0) at (2,0) {};
\node[vertex] (x3y0) at (3,0) {};
\node[vertex] (x4y0) at (4,0) {};
\node[vertex] (x0y1) at (0,1) {};
\node[vertex] (x1y1) at (1,1) {};
\node[vertex] (x2y1) at (2,1) {};
\node[vertex] (x3y1) at (3,1) {};
\node[vertex] (x0y2) at (0,2) {};
\node[vertex,use] (x1y2) at (1,2) {};
\node[vertex] (x2y2) at (2,2) {};
\node[vertex] (x0y3) at (0,3) {};
\node[vertex] (x1y3) at (1,3) {};
\node[vertex] (x0y4) at (0,4) {};

\node[anchor=center] at (x0y1.center) {\scriptsize\mbox{$\begin{array}{c}m^1_0(1), \\ 0\end{array}$}};
\node[anchor=center] at (x0y2.center) {\scriptsize\mbox{$\begin{array}{c}m^1_0(2), \\ 0\end{array}$}};
\node[anchor=center] at (x0y3.center) {\scriptsize\mbox{$\begin{array}{c}m^1_0(3), \\ 0\end{array}$}};

\node[anchor=center] at (x1y3.center) {\scriptsize\mbox{$\begin{array}{c}m^2_0(1), \\ 0\end{array}$}};
\node[anchor=center] at (x2y2.center) {\scriptsize\mbox{$\begin{array}{c}m^2_0(2), \\ 0\end{array}$}};
\node[anchor=center] at (x3y1.center) {\scriptsize\mbox{$\begin{array}{c}m^2_0(3), \\ 0\end{array}$}};

\node[anchor=center] at (x3y0.center) {\scriptsize\mbox{$\begin{array}{c}m^3_0(1), \\ 0\end{array}$}};
\node[anchor=center] at (x2y0.center) {\scriptsize\mbox{$\begin{array}{c}m^3_0(2), \\ 0\end{array}$}};
\node[anchor=center] at (x1y0.center) {\scriptsize\mbox{$\begin{array}{c}m^3_0(3), \\ 0\end{array}$}};

\node[anchor=center] at (x1y1.center) {\scriptsize\mbox{$\begin{array}{c}0 \end{array}$}};
\node[anchor=center] at (x2y1.center) {\scriptsize\mbox{$\begin{array}{c}0 \end{array}$}};
\node[anchor=center] at (x1y2.center) {\scriptsize\mbox{$\begin{array}{c}0 \end{array}$}};

\end{tikzpicture}
\endpgfgraphicnamed
}
\hspace{5mm}
\subfloat[$t=1$]{
\beginpgfgraphicnamed{pgffighext1}
\begin{tikzpicture}[scale=1,x={(15mm,0cm)},y={(7.5mm,15mm)}]
\tikzstyle{vertex}=[draw, rounded corners=2pt, minimum width=13mm, minimum height=13mm]
\tikzstyle{use}=[very thick,loosely dashed]
\setlength{\extrarowheight}{2.5pt}
\setcounter{tikzA}{4}

\node[vertex] (x0y0) at (0,0) {};
\node[vertex] (x1y0) at (1,0) {};
\node[vertex] (x2y0) at (2,0) {};
\node[vertex] (x3y0) at (3,0) {};
\node[vertex] (x4y0) at (4,0) {};
\node[vertex] (x0y1) at (0,1) {};
\node[vertex,use] (x1y1) at (1,1) {};
\node[vertex] (x2y1) at (2,1) {};
\node[vertex] (x3y1) at (3,1) {};
\node[vertex] (x0y2) at (0,2) {};
\node[vertex] (x1y2) at (1,2) {};
\node[vertex,use] (x2y2) at (2,2) {};
\node[vertex,use] (x0y3) at (0,3) {};
\node[vertex] (x1y3) at (1,3) {};
\node[vertex] (x0y4) at (0,4) {};

\node[anchor=center] at (x0y1.center) {\scriptsize\mbox{$\begin{array}{c}m^1_1(1), \\ m^3_0(3)\end{array}$}};
\node[anchor=center] at (x0y2.center) {\scriptsize\mbox{$\begin{array}{c}m^1_1(2), \\ 0\end{array}$}};
\node[anchor=center] at (x0y3.center) {\scriptsize\mbox{$\begin{array}{c}m^1_1(3), \\ 0\end{array}$}};

\node[anchor=center] at (x1y3.center) {\scriptsize\mbox{$\begin{array}{c}m^2_1(1), \\ m^1_0(3)\end{array}$}};
\node[anchor=center] at (x2y2.center) {\scriptsize\mbox{$\begin{array}{c}m^2_1(2), \\ 0\end{array}$}};
\node[anchor=center] at (x3y1.center) {\scriptsize\mbox{$\begin{array}{c}m^2_1(3), \\ 0\end{array}$}};

\node[anchor=center] at (x3y0.center) {\scriptsize\mbox{$\begin{array}{c}m^3_1(1), \\ m^2_0(3)\end{array}$}};
\node[anchor=center] at (x2y0.center) {\scriptsize\mbox{$\begin{array}{c}m^3_1(2), \\ 0\end{array}$}};
\node[anchor=center] at (x1y0.center) {\scriptsize\mbox{$\begin{array}{c}m^3_1(3), \\ 0\end{array}$}};

\node[anchor=center] at (x1y1.center) {\scriptsize\mbox{$\begin{array}{c}m^1_0(1) +\\ m^3_0(2) \end{array}$}};
\node[anchor=center] at (x2y1.center) {\scriptsize\mbox{$\begin{array}{c}m^2_0(2) +\\ m^3_0(1) \end{array}$}};
\node[anchor=center] at (x1y2.center) {\scriptsize\mbox{$\begin{array}{c}m^1_0(2) +\\ m^2_0(1) \end{array}$}};
\end{tikzpicture}
\endpgfgraphicnamed
}
\\
\subfloat[$t=2$]{
\beginpgfgraphicnamed{pgffighext2}
\begin{tikzpicture}[scale=1,x={(15mm,0cm)},y={(7.5mm,15mm)}]
\tikzstyle{vertex}=[draw, rounded corners=2pt, minimum width=13mm, minimum height=13mm]
\tikzstyle{use}=[very thick,loosely dashed]
\setlength{\extrarowheight}{2.5pt}
\setcounter{tikzA}{4}

\node[vertex] (x0y0) at (0,0) {};
\node[vertex] (x1y0) at (1,0) {};
\node[vertex] (x2y0) at (2,0) {};
\node[vertex] (x3y0) at (3,0) {};
\node[vertex] (x4y0) at (4,0) {};
\node[vertex] (x0y1) at (0,1) {};
\node[vertex] (x1y1) at (1,1) {};
\node[vertex,use] (x2y1) at (2,1) {};
\node[vertex] (x3y1) at (3,1) {};
\node[vertex,use] (x0y2) at (0,2) {};
\node[vertex] (x1y2) at (1,2) {};
\node[vertex] (x2y2) at (2,2) {};
\node[vertex] (x0y3) at (0,3) {};
\node[vertex,use] (x1y3) at (1,3) {};
\node[vertex] (x0y4) at (0,4) {};

\node[anchor=center] at (x0y1.center) {\scriptsize\mbox{$\begin{array}{c}m^1_2(1), \\ m^3_1(3)\end{array}$}};
\node[anchor=center] at (x0y2.center) {\scriptsize\mbox{$\begin{array}{c}m^1_2(2), \\ m^3_0(2)\end{array}$}};
\node[anchor=center] at (x0y3.center) {\scriptsize\mbox{$\begin{array}{c}m^1_2(3), \\ 0\end{array}$}};

\node[anchor=center] at (x1y3.center) {\scriptsize\mbox{$\begin{array}{c}m^2_2(1), \\ m^1_1(3)\end{array}$}};
\node[anchor=center] at (x2y2.center) {\scriptsize\mbox{$\begin{array}{c}m^2_2(2), \\ m^1_0(2)\end{array}$}};
\node[anchor=center] at (x3y1.center) {\scriptsize\mbox{$\begin{array}{c}m^2_2(3), \\ 0\end{array}$}};

\node[anchor=center] at (x3y0.center) {\scriptsize\mbox{$\begin{array}{c}m^3_2(1), \\ m^2_1(3)\end{array}$}};
\node[anchor=center] at (x2y0.center) {\scriptsize\mbox{$\begin{array}{c}m^3_2(2), \\ m^2_0(2)\end{array}$}};
\node[anchor=center] at (x1y0.center) {\scriptsize\mbox{$\begin{array}{c}m^3_2(3), \\ 0\end{array}$}};

\node[anchor=center] at (x1y1.center) {\scriptsize\mbox{$\begin{array}{c}m^1_1(1) +\\ m^2_0(1)+\\ m^3_1(2) \end{array}$}};
\node[anchor=center] at (x2y1.center) {\scriptsize\mbox{$\begin{array}{c}m^1_0(1) +\\ m^2_1(2) +\\ m^3_1(1) \end{array}$}};
\node[anchor=center] at (x1y2.center) {\scriptsize\mbox{$\begin{array}{c}m^1_1(2) +\\ m^2_1(1) +\\ m^3_0(1) \end{array}$}};
\end{tikzpicture}
\endpgfgraphicnamed
}
\hspace{5mm}
\subfloat[$t=3$]{
\beginpgfgraphicnamed{pgffighext3}
\begin{tikzpicture}[scale=1,x={(15mm,0cm)},y={(7.5mm,15mm)}]
\tikzstyle{vertex}=[draw, rounded corners=2pt, minimum width=13mm, minimum height=13mm]
\tikzstyle{result}=[very thick,dotted]
\setlength{\extrarowheight}{2.5pt}
\setcounter{tikzA}{4}

\node[vertex] (x0y0) at (0,0) {};
\node[vertex] (x1y0) at (1,0) {};
\node[vertex] (x2y0) at (2,0) {};
\node[vertex] (x3y0) at (3,0) {};
\node[vertex] (x4y0) at (4,0) {};
\node[vertex] (x0y1) at (0,1) {};
\node[vertex] (x1y1) at (1,1) {};
\node[vertex] (x2y1) at (2,1) {};
\node[vertex] (x3y1) at (3,1) {};
\node[vertex] (x0y2) at (0,2) {};
\node[vertex,result] (x1y2) at (1,2) {};
\node[vertex] (x2y2) at (2,2) {};
\node[vertex] (x0y3) at (0,3) {};
\node[vertex] (x1y3) at (1,3) {};
\node[vertex] (x0y4) at (0,4) {};

\node[anchor=center] at (x0y1.center) {\scriptsize\mbox{$\begin{array}{c}m^1_3(1), \\ m^3_2(3)\end{array}$}};
\node[anchor=center] at (x0y2.center) {\scriptsize\mbox{$\begin{array}{c}m^1_3(2), \\ m^3_1(2)\end{array}$}};
\node[anchor=center] at (x0y3.center) {\scriptsize\mbox{$\begin{array}{c}m^1_3(3), \\ m^3_0(1)\end{array}$}};

\node[anchor=center] at (x1y3.center) {\scriptsize\mbox{$\begin{array}{c}m^2_3(1), \\ m^1_2(3)\end{array}$}};
\node[anchor=center] at (x2y2.center) {\scriptsize\mbox{$\begin{array}{c}m^2_3(2), \\ m^1_1(2)\end{array}$}};
\node[anchor=center] at (x3y1.center) {\scriptsize\mbox{$\begin{array}{c}m^2_3(3), \\ m^1_0(1)\end{array}$}};

\node[anchor=center] at (x3y0.center) {\scriptsize\mbox{$\begin{array}{c}m^3_3(1), \\ m^2_2(3)\end{array}$}};
\node[anchor=center] at (x2y0.center) {\scriptsize\mbox{$\begin{array}{c}m^3_3(2), \\ m^2_1(2)\end{array}$}};
\node[anchor=center] at (x1y0.center) {\scriptsize\mbox{$\begin{array}{c}m^3_3(3), \\ m^2_0(1)\end{array}$}};

\node[anchor=center] at (x1y1.center) {\scriptsize\mbox{$\begin{array}{c}m^1_2(1) +\\ m^2_1(1)+\\ m^3_2(2) \end{array}$}};
\node[anchor=center] at (x2y1.center) {\scriptsize\mbox{$\begin{array}{c}m^1_1(1) +\\ m^2_2(2) +\\ m^3_2(1) \end{array}$}};
\node[anchor=center] at (x1y2.center) {\scriptsize\mbox{$\begin{array}{c}m^1_2(2) +\\ m^2_2(1) +\\ m^3_1(1) \end{array}$}};
\end{tikzpicture}
\endpgfgraphicnamed
}
\caption{Example operation of the network code of Section~\ref{sec:hex}, with $K=4$. The transmisssions of all nodes in the time slots $0,\dots,3$ are depicted. Different transmissions by the same node are separated by a comma. Note, that the symbol transmitted at $t=3$ by the node with dotted border can be obtained by summing, \ie taking the XOR of, all transmissions from nodes with a dashed border in earlier time slots. All nodes in the interior of the network perform this simple coding operation. \label{fig:hexexample}}
\end{figure*}
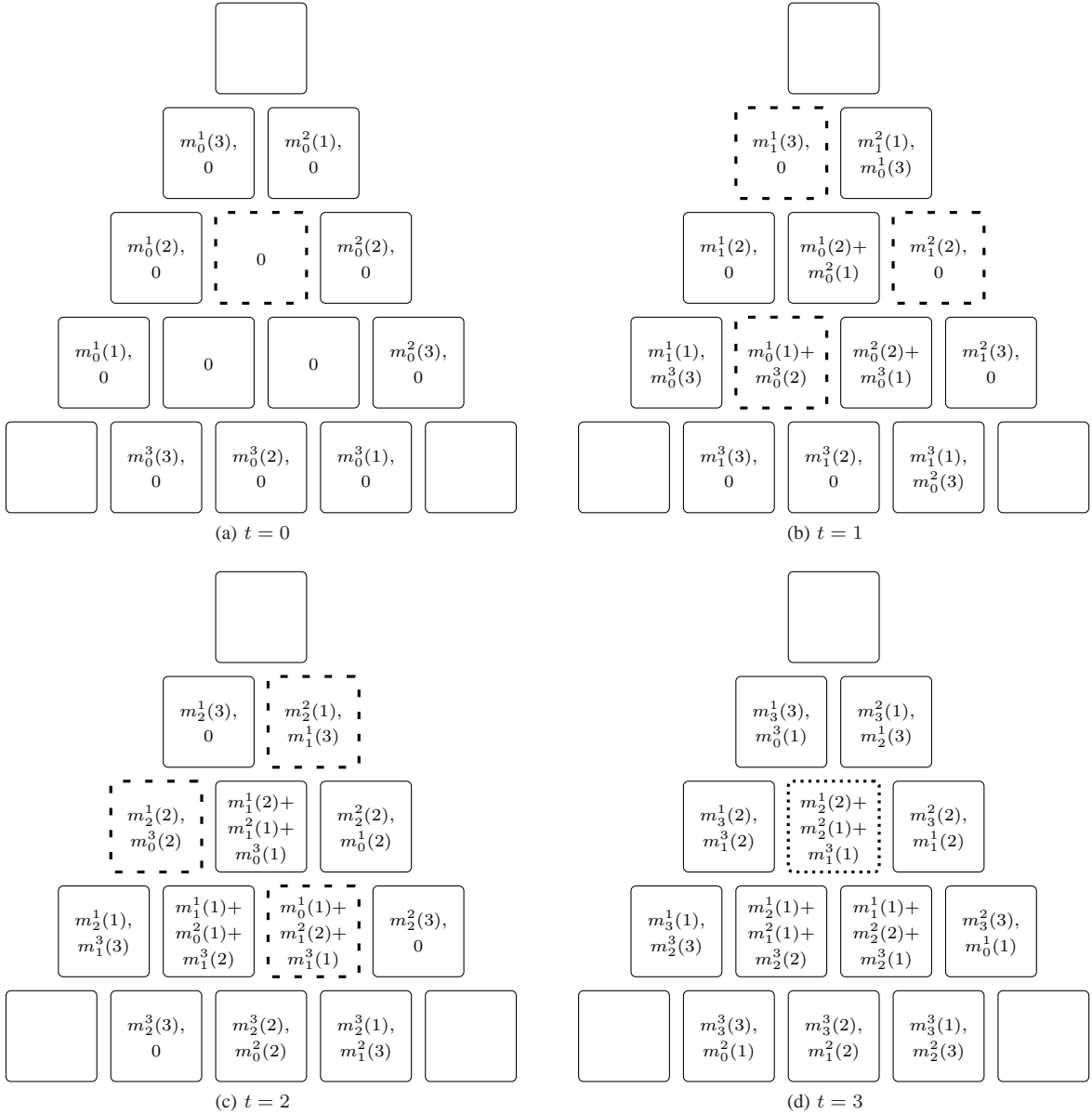
\subsection{Network Code} \label{sec:hexconstr}
The network code is such that in each time slot a new source symbol from each session is transmitted. Also, one symbol of each session is decoded by its destination in each time slot. After successfully decoding a symbol it is retransmitted by the destination in the next time slot. Nodes at the border will, therefore, transmit twice in each time slot. Nodes in the interior of the network transmit only once. The symbol that they transmit is a linear combination of one symbol from each of the sessions for which the shortest path between source and destination includes that node.

The operation of the network code is demonstrated in Figure~\ref{fig:hexexample} in which the transmissions of all nodes in the first four time slots are depicted. Different transmissions by the same node are separated by a comma. Note, moreover, that there is a startup phase, time slots $0$ to $2$, in which not all destinations are able to decode a symbol. From time slot $3$ onwards all destinations decode one symbol in every time slot. In analyzing the energy consumption of the coding scheme, we will ignore the startup phase.

The symbol transmitted at $t=3$ by the node with the dotted border can be obtained by summing, \ie taking the XOR of, all transmissions from nodes with a dashed border in earlier time slots. Indeed
\begin{multline} \label{eq:hexcodingopexample}
	m^1_1(3)+m^2_1(2)+m^1_0(1)+m^3_0(2)+m^2_2(1)+ \\
m^1_1(3)+m^1_2(2)+ m^3_0(2)+ m^1_0(1)+m^2_1(2)+m^3_1(1)= \\
m^1_2(2)+m^2_2(1)+m^3_1(1).
\end{multline}
This coding operation (\ie in time slot $t$ a node transmits the sum of what was transmitted by its top-left neighbour in time slot $t-2$, by its top right-neighbour in time slot $t-1$, etc., as visualized in Fig.~\ref{fig:hexexample}) is performed by all nodes that are in the interior of the network. The idea behind the coding operation is to cancel, by means of the XOR operation, all symbols that should not be retransmitted. In~\eqref{eq:hexcodingopexample}, for instance, we have $m^1_1(3)+m^1_1(3)=0$. The exact operation of the network code is made more precise in the remainder of this subsection. The coding operation for interior nodes is given in exact form in~\eqref{eq:hexcodingint}.

Nodes at the border of the network operate as follows. Let $0<u_2<K$. In time slot $t$ node $(0,u_2)$ transmits two symbols $x^1_t(0,u_2)$ and $x^3_t(0,u_2)$, where
\begin{empheq}[left=\text{Left border:}\quad,innerbox=\fbox]{align}
	&\ x^1_t(0,u_2) = m^1_t(u_2), \label{eq:hexcodinglefta} \\
	&\ x^3_t(0,u_2) = m^3_{t-u_2}(K-u_2). \label{eq:hexcodingleftretr}
\end{empheq}
Since $(0,u_2)$ is the source of session $m^1(u_2)$ it has source symbol $m^1_t(u_2)$ available. Also, $(0,u_2)$ is the destination for session $m^3(K-u_2)$. It remains to be shown that symbol $m^3_{t-u_2}(K-u_2)$ can be decoded by $(0,u_2)$ using the information obtained from its neighbours up to time slot $t$. For notational convenience let 
\begin{empheq}[left=\text{Left border:}\quad,innerbox=\fbox]{equation}
x_t(0,u_2)\triangleq x^1_t(0,u_2) +  x^3_t(0,u_2).
\end{empheq}
In similar fashion we have the following transmissions at the right and bottom borders of the network
\begin{empheq}[left=\text{Right border:}\quad,innerbox=\fbox]{align}
	x^1_t(v_1,v_2) &= m^1_{t-v_1}(v_2), \label{eq:hexcodingrighta} \\
	x^2_t(v_1,v_2) &= m^2_{t}(v_1), \label{eq:hexcodingrightb} \\ 
	x_t(v_1,v_2) &\triangleq x^1_t(v_1,v_2) +  x^2_t(v_1,v_2),
\end{empheq}
\begin{empheq}[left=\text{Bottom border:}\quad,innerbox=\fbox]{align}
	x^2_t(u_1,0) &= m^2_{t-K+u_1}(u_1), \label{eq:hexcodingbottomretr} \\
	x^3_t(u_1,0) &= m^3_{t}(K-u_1),  \\
	x_t(u_1,0) &\triangleq x^2_t(u_1,0) +  x^3_t(u_1,0),
\end{empheq}
where $u_1,v_1,v_2>0$, $u_1,v_1,v_2<K$ and $v_1+v_2=K$. Moreover, $x_t(v_1,v_2)$ and $x_t(u_1,0)$ are not symbols that are transmitted, but only notational shortcuts.

Nodes in the interior of the network transmit once in each time slot. Let $(u_1,u_2)\in\VVint$. The coding operation it performs is given by
\begin{empheq}[box=\fbox]{align} \label{eq:hexcodingint}
	x_t(u_1, & u_2) = x_{t-1}(u_1-1,u_2) + \notag \\
&\ x_{t-2}(u_1-1,u_2+1) + x_{t-1}(u_1,u_2+1) + \notag \\
&\ x_{t-3}(u_1,u_2) + x_{t-2}(u_1+1,u_2) + \notag \\
&\ x_{t-2}(u_1,u_2-1) + x_{t-1}(u_1+1,u_2-1).
\end{empheq}

\subsection{Validity of the Network Code} \label{sec:hexvalid}
We need to show that destinations can decode in time in order to retransmit the required symbols according to~(\ref{eq:hexcodingleftretr}), (\ref{eq:hexcodingrighta}) and~(\ref{eq:hexcodingbottomretr}). In order to do so we first analyze how data propagates through the network. If we look at the nodes in the network that transmit linear combinations that contain a certain source symbol, we see that symbols propagate exactly along the shortest paths between source and destination. This is made more precise in the following two lemmas.
\begin{lemma} \label{lem:hexsinglesymbol}
Let $0<u_2<K$. Assume that the only non-zero source symbol transmitted in the network is $m_0^1(u_2)$ by node $(0,u_2)$ in time slot $0$. Then, for all $t\geq 0$ and $(v_1,v_2)\in\VVint$
\begin{equation}
x_t(v_1,v_2) = 
\begin{cases}
	m_0^1(u_2),\quad &\text{if } v_1=t, v_2=u_2, \\
	0,\quad &\text{otherwise.}
\end{cases}
\end{equation}
\end{lemma}
\begin{IEEEproof}
We use induction over time. The base case is time slot $t=0$, for which it is readilly verified that the statement is true. Now, for the induction step suppose that the lemma holds for all $t'$ smaller than $t$. This implies that for all $\tau>0$ and $(v_1,v_2)\in\VVint$,
\begin{equation}
	x_{t-\tau}(v_1,v_2) = x_{t-\tau-1}(v_1-1,v_2).
\end{equation}
Hence,
\begin{align}
	x_t(v_1,v_2) =
&\ x_{t-1}(v_1-1,v_2) + \notag \\
&\ x_{t-2}(v_1-1,v_2+1) + x_{t-1}(v_1,v_2+1) + \notag \\
&\ x_{t-3}(v_1,v_2) + x_{t-2}(v_1+1,v_2) + \notag \\
&\ x_{t-2}(v_1,v_2-1) + x_{t-1}(v_1+1,v_2-1) \notag \\
= &\ x_{t-1}(v_1-1,v_2)+ \notag \\
&\ x_{t-2}(v_1-1,v_2+1) + x_{t-2}(v_1-1,v_2+1) + \notag \\
&\ x_{t-3}(v_1,v_2) + x_{t-3}(v_1,v_2) + \notag \\
&\ x_{t-2}(v_1,v_2-1) + x_{t-2}(v_1,v_2-1) \notag \\
= &\ x_{t-1}(v_1-1,v_2),
\end{align}
which by the induction hypothesis is equal to $m_0^1(u_2)$ if $v_1=t$ and $v_2=u_2$ and zero otherwise.
\end{IEEEproof}

\begin{lemma} \label{lem:hexform}
Let $(u_1,u_2)\in\VVint$.
\begin{multline*}
x_t(u_1,u_2) = m_{t-u_1}^1(u_2) + \\ m_{t-K+u_1+u_2}^2(u_1) + m_{t-u_2}^3(K-u_1-u_2).
\end{multline*}
\end{lemma}
\begin{IEEEproof}
From Lemma~\ref{lem:hexsinglesymbol}, the time-invariance of the system, and the symmetry of the coding operation~(\ref{eq:hexcodingint}) of the internal nodes.
\end{IEEEproof}

We are now ready to prove that the destinations can correctly decode source symbols. We present the decoding procedure for nodes on the right border of the network. The decoding procedures at the other borders can be obtained by exploiting the symmetry of the system.
\begin{lemma}
Consider $(u_1,u_2)$, with $u_1+u_2=K$, $0<u_2<K$, \ie the destination of session $m^1(u_2)$. It can decode symbol $m_{t-u_1}^1(u_2)$ at the end of time slot $t-1$ as
\begin{multline} \label{eq:hexdec}
x^2_{t-2}(u_1-1,u_2+1) + x_{t-1}(u_1-1,u_2) + x^2_{t-3}(u_1,u_2) + \\
x_{t-2}(u_1,u_2-1) + x^1_{t-1}(u_1+1,u_2-1).
\end{multline}
\end{lemma}
\begin{IEEEproof}
From Lemma~\ref{lem:hexform}, (\ref{eq:hexcodingrighta}) and (\ref{eq:hexcodingrightb}) it follows that~(\ref{eq:hexdec}) equals
\begin{multline}
m^1_{t-u_1}(u_2) + m^1_{t-u_1-2}(u_2-1) + m^1_{t-u_1-2}(u_2-1) + \\
m^2_{t-2}(u_1-1) + m^2_{t-2}(u_1-1) +
m^2_{t-3}(u_1) + m^2_{t-3}(u_1) +\\
m^3_{t-u_2-1}(1) + m^3_{t-u_2-1}(1) \\
= m^1_{t-u_1}(u_2).
\end{multline}
\end{IEEEproof}

\subsection{Energy Consumption} \label{sec:hexconcl}
The energy consumption of the network coding scheme presented above is given in the following lemma. 
\begin{lemma} \label{lem:hexenc}
$\min_r\Enc(V,M,r)\leq \Enc(V,M,1)\leq \frac{c}{2} K^2 + \OO(K)$.
\end{lemma}
\begin{IEEEproof}
From~(\ref{eq:hexcodinglefta})--(\ref{eq:hexcodingint}) we have that each of the $3(K-1)$ nodes at the border that are source or destination are transmitting twice in each time slot. Each of the $(K-1)(K-2)/2$ internal nodes is transmitting once in each time slot. Since $r=1$, the energy consumption per transmission is $c$. This gives
\begin{multline}
\Enc(V,M,1) \leq 6c(K-1) + c(K-1)(K-2)/2 \\ = \frac{c}{2}K^2 + \OO(K).
\end{multline}
\end{IEEEproof}

Next, we give the minimum energy required by a routing solution.
\begin{lemma} \label{lem:hexer}
$\min_r\Er(V,M,r) = \Er(V,M,1) = \frac{3 c}{2} K^2 + \OO(K)$.
\end{lemma}
\begin{IEEEproof}
Since we consider routing we need to take the shortest path for each session. Since the energy consumption per hop equals $c r^\alpha$, the energy consumption under routing is minimized for $r=1$. Now, we see that the number of transmissions required to deliver a symbol for the sessions $m^1(1),\dots,m^1(K-1)$ equals $K(K-1)/2$. Adding the transmissions for sessions of type $2$ and $3$ gives
\begin{equation}
\Er(V,M,1) =  \frac{3 c}{2} K(K-1) = \frac{3 c}{2} K^2 + \OO(K).
\end{equation}
\end{IEEEproof}

Using the above two lemmas we are able to prove Theorem~\ref{th:var}.
\begin{IEEEproof}[Proof of Theorem~\ref{th:var}]
Remember, that $\Bvar$ is defined as the maximum of $\min_r \Er(V,M,r)/\min_r \Enc(V,M,r)$ over $V$ and $M$. Hence, $\min_r \Er(V,M,r)/\min_r \Enc(V,M,r)$ for any specific $V$ and $M$ will provide a lower bound to $\Bvar$. In addition, any upper bound to $\min_r \Enc(V,M,r)$ will result in a lower bound to $\Bvar$. Hence, from Lemmas~\ref{lem:hexenc} and~\ref{lem:hexer} we have
\begin{align}
\Bvar(2)
&\geq \lim_{K\to\infty} \frac{\min_r \Er(V,M,r)}{\min_r \Enc(V,M,r)} \notag \\
&\geq \lim_{K\to\infty} \frac{\Er(V,M,1)}{\Enc(V,M,1)} \notag \\
&\geq \lim_{K\to\infty} \frac{\frac{3 c}{2} K^2 + \OO(K)}{\frac{c}{2} K^2 + \OO(K)} \notag \\
& = 3.
\end{align}
\end{IEEEproof}

%
%
%
\section{An Efficient Code on the $d$-dimensional Rectangular Lattice} \label{sec:rect}
In this section we present a multiple unicast configuration in which the nodes are placed at integer coordinates in a $d$-dimensional space, \ie at the rectangular lattice.

\begin{figure}
\begin{center}
\beginpgfgraphicnamed{pgffigrectlattice}
\begin{tikzpicture}[scale=1]
\tikzstyle{vertex}=[circle,fill, minimum size=8pt, inner sep=0pt]

\setcounter{tikzA}{5}

\foreach \x in {0,...,\thetikzA}
	\foreach \y in {0,...,\thetikzA}
		\node[vertex] (x\x y\y) at (\x,\y) {};

\setcounter{tikzB}{\thetikzA-1}
\foreach \x in {0,...,\thetikzB}
	{
	\setcounter{tikzC}{\x+1}
	\draw (x\x y0) to (x\thetikzC y0);
	\draw (x\x y0) to (x\thetikzC y1);
	\draw (x\x y0) to (x\x y1);
	\foreach \y in {1,...,\thetikzB}
		{
		\setcounter{tikzD}{\y-1}
		\setcounter{tikzE}{\y+1}
		\draw (x\x y\y) to (x\thetikzC y\thetikzD);
		\draw (x\x y\y) to (x\thetikzC y\y);
		\draw (x\x y\y) to (x\thetikzC y\thetikzE);
		\draw (x\x y\y) to (x\x y\thetikzE);
		}
	\draw (x\x y\thetikzA) to (x\thetikzC y\thetikzA);
	\draw (x\x y\thetikzA) to (x\thetikzC y\thetikzB);
	\draw (x\thetikzA y\x) to (x\thetikzA y\thetikzC);
	}

\node[anchor=east] at (x0y0.west) {\scriptsize\mbox{$(0,0)$}};
\node[anchor=east] at (x0y1.west) {\scriptsize\mbox{$(0,1)$}};
\node[anchor=east] at (x0y\thetikzA.west) {\scriptsize\mbox{$(0,K)$}};

\node[anchor=north] at (x1y0.south) {\scriptsize\mbox{$(1,0)$}};
\node[anchor=north] at (x\thetikzA y0.south) {\scriptsize\mbox{$(K,0)$}};

\end{tikzpicture}
\endpgfgraphicnamed
\caption{Nodes at a subset of the $d$-dimensional rectangular lattice, $d=2$ depicted in the figure, with the connectivity induced by a transmission range $r=\sqrt{d}$. The size of the network is controlled by $K$, with $K=5$ in this figure. \label{fig:rectlattice}}
\end{center}

\begin{center}
\beginpgfgraphicnamed{pgffigrectsessions}
\begin{tikzpicture}[scale=1]
\tikzstyle{vertex}=[circle,fill, minimum size=8pt, inner sep=0pt]

\setcounter{tikzA}{5}

\foreach \x in {0,...,\thetikzA}
	\foreach \y in {0,...,\thetikzA}
		\node[vertex] (x\x y\y) at (\x,\y) {};

\setcounter{tikzB}{\thetikzA-1}
\foreach \y in {1,...,\thetikzB}
	{
	\draw[<->] (x0y\y) to (x\thetikzA y\y);
	\draw[<->] (x\y y0) to (x\y y\thetikzA);
	\node[anchor=east] at (x0y\y.west) {\scriptsize\mbox{$s^1(\y), r^3(\y)$}};
	\node[anchor=west] at (x\thetikzA y\y.east) {\scriptsize\mbox{$s^3(\y), r^1(\y)$}};
	\node[anchor=north] at (x\y y0.south) {\scriptsize\mbox{$\begin{array}{c}s^2(\y)\\ r^4(\y)\end{array}$}};
	\node[anchor=south] at (x\y y\thetikzA.north) {\scriptsize\mbox{$\begin{array}{c}s^4(\y)\\ r^2(\y)\end{array}$}};
	}

\end{tikzpicture}
\endpgfgraphicnamed
\caption{The multiple unicast sessions on the network from Figure~\ref{fig:rectlattice}. \label{fig:rectsessions}}
\end{center}
\end{figure}

%
\subsection{Configuration} \label{sec:rectconf}
The size of the configuration is parametrized by a positive integer $K$. We have
\begin{equation}
V = \{ (v_1,\dots,v_d) \mid 0\leq v_i\leq K, i=1,\dots,d\}.
\end{equation}
The interior of the network is given by
\begin{equation}
\VVint = \{ v\in V \mid 0<v_i<K, i=1,\dots,d\}.
\end{equation}
We will make us of
\begin{equation}
\bar V = \{ v\in V \mid \exists\text{ unique } i: v_i\in\{0,K\} \},
\end{equation}
which corresponds to those nodes that are part of exactly one face of the network.

The transmission range that will be used is $r=\sqrt{d}$. This transmission range induces a neighbourhood consisting of all neighbours within distance $\sqrt{d}$. The coding operation of our network code is based on only part of the neighbourhood, \ie it uses
\begin{equation}
	\NN_\vv=\{\uu\in\VV \mid |u_i-v_i|\leq 1\ \forall i, \uu\neq\vv\}.
\end{equation}
Note, that for $d\leq 3$, $\NN_\vv$ corresponds to the complete neighbourhood of $v$. We will be using $\dist(\uu,\vv)\triangleq\|\uu-\vv\|_1=\sum_{i=1}^d|u_i-v_i|$, \ie $\dist(\uu,\vv)$ denotes the Manhattan distance from $\uu$ to $\vv$. The network and its connectivity are depicted for $d=2$ in Figure~\ref{fig:rectlattice}.

A source is located at each $v\in\bar V$. Therefore, there are $|\bar V|=2d(K-1)^{d-1}$ sessions. If $v_i=0$, we denote the session corresponding to this source by $m^i(\vvm{i})$. Recall from Section~\ref{sec:model} that $\vvm{i}$ denotes the $d-1$ dimensional vector obtained by removing the $i$-th element from $\vv$. If $v_i=K$, we denote the session by $m^{d+i}(\vvm{i})$. The destination of each session is located at the other side of the network, \ie we have $r^i(\vvm{i})=s^{d+i}(\vvm{i})$ and $r^{d+i}(\vvm{i})=s^i(\vvm{i})$. The positions of sources and destinations are depicted for $d=2$ in Figure~\ref{fig:rectsessions}. It can be seen that $m^i(\vvm{i})$ and $m^{d+i}(\vvm{i})$ form oppositely directed sessions.

\subsection{Network code}
We introduce sets $\Theta_\delta\subset\{1,\dots,2d\}$, $0\leq\delta\leq d$, which are defined recursively as follows
\begin{align}
	\Theta_d &= \{d\}, \\
	\Theta_\delta &= (\Theta_{\delta+1}-1)\Delta(\Theta_{\delta+1}+1),\quad 0<\delta<d, \\
	\Theta_0 &= \big((\Theta_1-1)\Delta(\Theta_1+1)\big)\setminus\{0\},
\end{align}
where $\Delta$ denotes symmetric difference and $\Theta_\delta\pm 1 = \{\tau\pm 1 | \tau\in\Theta_\delta\}$.
Note, that irrespective of $d$ we have $1\in\Theta_1$. As an example for $d=2$ we have $\Theta_2=\{2\}$, $\Theta_1=\{1,3\}$ and $\Theta_0=\{4\}$.

\begin{figure*}
\centering
\subfloat[$t=0$]{
\beginpgfgraphicnamed{pgffigrectt0}
\begin{tikzpicture}[scale=1,x={(18mm,0cm)},y={(0,18mm)}]
\tikzstyle{vertex}=[draw, rounded corners=2pt, minimum width=16mm, minimum height=16mm]
\tikzstyle{use}=[very thick,loosely dashed]
\setlength{\extrarowheight}{2.5pt}
\setcounter{tikzA}{3}

\node[vertex] (x0y0) at (0,0) {};
\node[vertex] (x1y0) at (1,0) {};
\node[vertex] (x2y0) at (2,0) {};
\node[vertex] (x3y0) at (3,0) {};
\node[vertex] (x0y1) at (0,1) {};
\node[vertex,use] (x1y1) at (1,1) {};
\node[vertex] (x2y1) at (2,1) {};
\node[vertex] (x3y1) at (3,1) {};
\node[vertex,use] (x0y2) at (0,2) {};
\node[vertex] (x1y2) at (1,2) {};
\node[vertex,use] (x2y2) at (2,2) {};
\node[vertex] (x3y2) at (3,2) {};
\node[vertex] (x0y3) at (0,3) {};
\node[vertex,use] (x1y3) at (1,3) {};
\node[vertex] (x2y3) at (2,3) {};
\node[vertex] (x3y3) at (3,3) {};

\node[anchor=center] at (x0y1.center) {\scriptsize\mbox{$\begin{array}{c}m^1_0(1), \\ 0\end{array}$}};
\node[anchor=center] at (x0y2.center) {\scriptsize\mbox{$\begin{array}{c}m^1_0(2), \\ 0\end{array}$}};

\node[anchor=center] at (x1y0.center) {\scriptsize\mbox{$\begin{array}{c}m^2_0(1), \\ 0\end{array}$}};
\node[anchor=center] at (x2y0.center) {\scriptsize\mbox{$\begin{array}{c}m^2_0(2), \\ 0\end{array}$}};

\node[anchor=center] at (x3y1.center) {\scriptsize\mbox{$\begin{array}{c}m^3_0(1), \\ 0\end{array}$}};
\node[anchor=center] at (x3y2.center) {\scriptsize\mbox{$\begin{array}{c}m^3_0(2), \\ 0\end{array}$}};

\node[anchor=center] at (x1y3.center) {\scriptsize\mbox{$\begin{array}{c}m^4_0(1), \\ 0\end{array}$}};
\node[anchor=center] at (x2y3.center) {\scriptsize\mbox{$\begin{array}{c}m^4_0(2), \\ 0\end{array}$}};

\node[anchor=center] at (x1y1.center) {\scriptsize\mbox{$\begin{array}{c}0 \end{array}$}};
\node[anchor=center] at (x2y1.center) {\scriptsize\mbox{$\begin{array}{c}0 \end{array}$}};
\node[anchor=center] at (x1y2.center) {\scriptsize\mbox{$\begin{array}{c}0 \end{array}$}};
\node[anchor=center] at (x2y2.center) {\scriptsize\mbox{$\begin{array}{c}0 \end{array}$}};

\end{tikzpicture}
\endpgfgraphicnamed
}
\hspace{5mm}
\subfloat[$t=1$]{
\beginpgfgraphicnamed{pgffigrectt1}
\begin{tikzpicture}[scale=1,x={(18mm,0cm)},y={(0,18mm)}]
\tikzstyle{vertex}=[draw, rounded corners=2pt, minimum width=16mm, minimum height=16mm]
\tikzstyle{use}=[very thick,loosely dashed]
\setlength{\extrarowheight}{2.5pt}
\setcounter{tikzA}{3}

\node[vertex] (x0y0) at (0,0) {};
\node[vertex] (x1y0) at (1,0) {};
\node[vertex] (x2y0) at (2,0) {};
\node[vertex] (x3y0) at (3,0) {};
\node[vertex,use] (x0y1) at (0,1) {};
\node[vertex] (x1y1) at (1,1) {};
\node[vertex,use] (x2y1) at (2,1) {};
\node[vertex] (x3y1) at (3,1) {};
\node[vertex] (x0y2) at (0,2) {};
\node[vertex] (x1y2) at (1,2) {};
\node[vertex] (x2y2) at (2,2) {};
\node[vertex] (x3y2) at (3,2) {};
\node[vertex,use] (x0y3) at (0,3) {};
\node[vertex] (x1y3) at (1,3) {};
\node[vertex,use] (x2y3) at (2,3) {};
\node[vertex] (x3y3) at (3,3) {};

\node[anchor=center] at (x0y1.center) {\scriptsize\mbox{$\begin{array}{c}m^1_1(1), \\ 0\end{array}$}};
\node[anchor=center] at (x0y2.center) {\scriptsize\mbox{$\begin{array}{c}m^1_1(2), \\ 0\end{array}$}};

\node[anchor=center] at (x1y0.center) {\scriptsize\mbox{$\begin{array}{c}m^2_1(1), \\ 0\end{array}$}};
\node[anchor=center] at (x2y0.center) {\scriptsize\mbox{$\begin{array}{c}m^2_1(2), \\ 0\end{array}$}};

\node[anchor=center] at (x3y1.center) {\scriptsize\mbox{$\begin{array}{c}m^3_1(1), \\ 0\end{array}$}};
\node[anchor=center] at (x3y2.center) {\scriptsize\mbox{$\begin{array}{c}m^3_1(2), \\ 0\end{array}$}};

\node[anchor=center] at (x1y3.center) {\scriptsize\mbox{$\begin{array}{c}m^4_1(1), \\ 0\end{array}$}};
\node[anchor=center] at (x2y3.center) {\scriptsize\mbox{$\begin{array}{c}m^4_1(2), \\ 0\end{array}$}};

\node[anchor=center] at (x1y1.center) {\scriptsize\mbox{$\begin{array}{c}m^1_0(1) +\\ m^2_0(1) \end{array}$}};
\node[anchor=center] at (x2y1.center) {\scriptsize\mbox{$\begin{array}{c}m^2_0(2) +\\ m^3_0(1)\end{array}$}};
\node[anchor=center] at (x1y2.center) {\scriptsize\mbox{$\begin{array}{c}m^1_0(2) +\\ m^4_0(1) \end{array}$}};
\node[anchor=center] at (x2y2.center) {\scriptsize\mbox{$\begin{array}{c}m^3_0(2) +\\ m^4_0(2) \end{array}$}};
\end{tikzpicture}
\endpgfgraphicnamed
}
\\
\subfloat[$t=2$]{
\beginpgfgraphicnamed{pgffigrectt2}
\begin{tikzpicture}[scale=1,x={(18mm,0cm)},y={(0,18mm)}]
\tikzstyle{vertex}=[draw, rounded corners=2pt, minimum width=16mm, minimum height=16mm]
\tikzstyle{use}=[very thick,loosely dashed]
\setlength{\extrarowheight}{2.5pt}
\setcounter{tikzA}{3}

\node[vertex] (x0y0) at (0,0) {};
\node[vertex] (x1y0) at (1,0) {};
\node[vertex] (x2y0) at (2,0) {};
\node[vertex] (x3y0) at (3,0) {};
\node[vertex] (x0y1) at (0,1) {};
\node[vertex,use] (x1y1) at (1,1) {};
\node[vertex] (x2y1) at (2,1) {};
\node[vertex] (x3y1) at (3,1) {};
\node[vertex,use] (x0y2) at (0,2) {};
\node[vertex] (x1y2) at (1,2) {};
\node[vertex,use] (x2y2) at (2,2) {};
\node[vertex] (x3y2) at (3,2) {};
\node[vertex] (x0y3) at (0,3) {};
\node[vertex,use] (x1y3) at (1,3) {};
\node[vertex] (x2y3) at (2,3) {};
\node[vertex] (x3y3) at (3,3) {};

\node[anchor=center] at (x0y1.center) {\scriptsize\mbox{$\begin{array}{c}m^1_2(1), \\ 0\end{array}$}};
\node[anchor=center] at (x0y2.center) {\scriptsize\mbox{$\begin{array}{c}m^1_2(2), \\ 0\end{array}$}};

\node[anchor=center] at (x1y0.center) {\scriptsize\mbox{$\begin{array}{c}m^2_2(1), \\ 0\end{array}$}};
\node[anchor=center] at (x2y0.center) {\scriptsize\mbox{$\begin{array}{c}m^2_2(2), \\ 0\end{array}$}};

\node[anchor=center] at (x3y1.center) {\scriptsize\mbox{$\begin{array}{c}m^3_2(1), \\ 0\end{array}$}};
\node[anchor=center] at (x3y2.center) {\scriptsize\mbox{$\begin{array}{c}m^3_2(2), \\ 0\end{array}$}};

\node[anchor=center] at (x1y3.center) {\scriptsize\mbox{$\begin{array}{c}m^4_2(1), \\ 0\end{array}$}};
\node[anchor=center] at (x2y3.center) {\scriptsize\mbox{$\begin{array}{c}m^4_2(2), \\ 0\end{array}$}};

\node[anchor=center] at (x1y1.center) {\scriptsize\mbox{$\begin{array}{c}m^1_1(1) +\\ m^2_1(1) +\\ m^3_0(1) +\\ m^4_0(1) \end{array}$}};
\node[anchor=center] at (x2y1.center) {\scriptsize\mbox{$\begin{array}{c}m^1_0(1) +\\ m^2_1(2) +\\ m^3_1(1) +\\ m^4_0(2) \end{array}$}};
\node[anchor=center] at (x1y2.center) {\scriptsize\mbox{$\begin{array}{c}m^1_1(2) +\\ m^2_0(1) +\\ m^3_0(2) +\\ m^4_1(1) \end{array}$}};
\node[anchor=center] at (x2y2.center) {\scriptsize\mbox{$\begin{array}{c}m^1_0(2) +\\ m^2_0(2) +\\ m^3_1(2) +\\ m^4_1(2) \end{array}$}};
\end{tikzpicture}
\endpgfgraphicnamed
}
\hspace{5mm}
\subfloat[$t=3$]{
\beginpgfgraphicnamed{pgffigrectt3}
\begin{tikzpicture}[scale=1,x={(18mm,0cm)},y={(0,18mm)}]
\tikzstyle{vertex}=[draw, rounded corners=2pt, minimum width=16mm, minimum height=16mm]
\tikzstyle{result}=[very thick,dotted]
\setlength{\extrarowheight}{2.5pt}
\setcounter{tikzA}{3}

\node[vertex] (x0y0) at (0,0) {};
\node[vertex] (x1y0) at (1,0) {};
\node[vertex] (x2y0) at (2,0) {};
\node[vertex] (x3y0) at (3,0) {};
\node[vertex] (x0y1) at (0,1) {};
\node[vertex] (x1y1) at (1,1) {};
\node[vertex] (x2y1) at (2,1) {};
\node[vertex] (x3y1) at (3,1) {};
\node[vertex] (x0y2) at (0,2) {};
\node[vertex,result] (x1y2) at (1,2) {};
\node[vertex] (x2y2) at (2,2) {};
\node[vertex] (x3y2) at (3,2) {};
\node[vertex] (x0y3) at (0,3) {};
\node[vertex] (x1y3) at (1,3) {};
\node[vertex] (x2y3) at (2,3) {};
\node[vertex] (x3y3) at (3,3) {};

\node[anchor=center] at (x0y1.center) {\scriptsize\mbox{$\begin{array}{c}m^1_3(1), \\ m^3_0(1)\end{array}$}};
\node[anchor=center] at (x0y2.center) {\scriptsize\mbox{$\begin{array}{c}m^1_3(2), \\ m^3_0(2)\end{array}$}};

\node[anchor=center] at (x1y0.center) {\scriptsize\mbox{$\begin{array}{c}m^2_3(1), \\ m^4_0(1)\end{array}$}};
\node[anchor=center] at (x2y0.center) {\scriptsize\mbox{$\begin{array}{c}m^2_3(2), \\ m^4_0(2)\end{array}$}};

\node[anchor=center] at (x3y1.center) {\scriptsize\mbox{$\begin{array}{c}m^3_3(1), \\ m^1_0(1)\end{array}$}};
\node[anchor=center] at (x3y2.center) {\scriptsize\mbox{$\begin{array}{c}m^3_3(2), \\ m^1_0(2)\end{array}$}};

\node[anchor=center] at (x1y3.center) {\scriptsize\mbox{$\begin{array}{c}m^4_3(1), \\ m^2_0(1)\end{array}$}};
\node[anchor=center] at (x2y3.center) {\scriptsize\mbox{$\begin{array}{c}m^4_3(2), \\ m^2_0(2)\end{array}$}};

\node[anchor=center] at (x1y1.center) {\scriptsize\mbox{$\begin{array}{c}m^1_2(1) +\\ m^2_2(1) +\\ m^3_1(1) +\\ m^4_1(1) \end{array}$}};
\node[anchor=center] at (x2y1.center) {\scriptsize\mbox{$\begin{array}{c}m^1_1(1) +\\ m^2_2(2) +\\ m^3_2(1) +\\ m^4_1(2) \end{array}$}};
\node[anchor=center] at (x1y2.center) {\scriptsize\mbox{$\begin{array}{c}m^1_2(2) +\\ m^2_1(1) +\\ m^3_1(2) +\\ m^4_2(1) \end{array}$}};
\node[anchor=center] at (x2y2.center) {\scriptsize\mbox{$\begin{array}{c}m^1_1(2) +\\ m^2_1(2) +\\ m^3_2(2) +\\ m^4_2(2) \end{array}$}};
\end{tikzpicture}
\endpgfgraphicnamed
}
\caption{Example operation of the network code of Section~\ref{sec:rect}, with $K=3$. The transmisssions of all nodes in the time slots $0,\dots,3$ are depicted. Different transmissions by the same node are separated by a comma. Note, that the symbol transmitted at $t=3$ by the node with dotted border can be obtained by summing, \ie taking the XOR of, all transmissions from nodes with a dashed border in earlier time slots. All nodes in the interior of the network perform this simple coding operation. \label{fig:rectexample}}
\end{figure*}
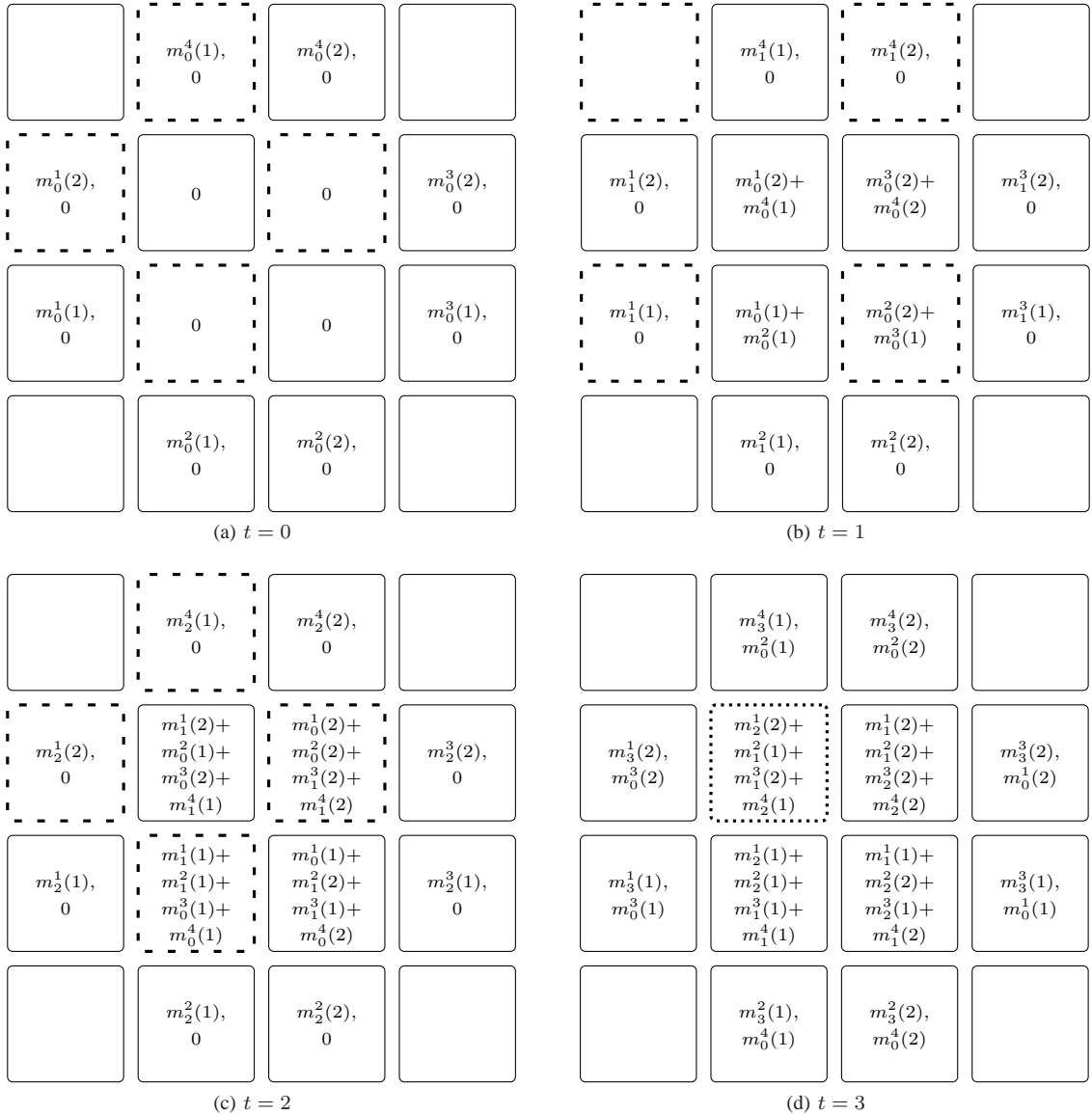
The scheme is very similar in flavour to the scheme presented in Section~\ref{sec:hex}, its operation is demonstrated in Figure~\ref{fig:rectexample} in which, for $d=2$ and $K=3$, the transmissions of all nodes in the first four time slots are depicted. The operation of the scheme is such that in time slot $t$ sources transmit the $t$-th source symbol and destinations decode the $(t-K)$-th source symbol. Besides transmitting a new source symbol in each time slot, sources/destinations will also retransmit the symbol that has been decoded in that time slot, \ie they transmit two different symbols in each time slot. In the figure, different transmissions by the same node are separated by a comma. Nodes in the interior of the network transmit only once. The symbol that they transmit is a linear combination of one symbol from each of the sessions for which the shortest path between source and destination includes that node. The symbol transmitted at $t=3$ by the node with the dotted border can be obtained by summing, \ie taking the XOR of, all transmissions from nodes with a dashed border in earlier time slots. This coding operation is performed by all nodes that are in the interior of the network. The exact operation of the network code is made more precise in the remainder of this subsection. The coding operation for interior nodes is given in exact form in~\eqref{eq:rectcodingint}.

Let node $\vv\in\bar V$. Remember, that $\vv\in\bar V$ implies that there exists a unique $i$ such that $v_i\in\{0,K\}$. Node $\vv$ transmits 
\begin{equation} \label{eq:rectcoding1} \boxed{
x^i_t(\vv) = m^i_{t-v_i}(\vvm{i})
}\end{equation}
and
\begin{equation} \label{eq:rectcoding2} \boxed{
x^{d+i}_t(\vv) = m^{d+i}_{t-K+v_i}(\vvm{i}).
}\end{equation}
For notational convenience, let
\begin{equation} \label{eq:rectcoding3} \boxed{
	x_t(\vv) \triangleq x^i_t(\vv) + x^{d+i}_t(\vv).
}\end{equation}
The coding operation performed by an internal node is as follows
\begin{equation} \label{eq:rectcodingint} \boxed{
	x_t(\vv) = \sum_{\uu\in\NN_\vv\cup\{\vv\}}\sum_{\tau\in\Theta_{\dist(\uu,\vv)}}
	x_{t-\tau}(\uu).}
\end{equation}

\subsection{Validity of the Network Code}

The following result follows directly from the definition of the sets $\Theta_\delta$, but is stated here as a lemma because of its importance in the remainder of the paper.
\begin{lemma} \label{lem:recttheta}
Let $\{x_t\}$ be a sequence of symbols from $\mathbb{F}_2$ and let $0<\delta<d$. We have
\begin{align}
	\sum_{\tau\in\Theta_{\delta}} x_{t-\tau} &= \sum_{\tau\in\Theta_{\delta+1}} \left[ x_{t-\tau+1} + x_{t-\tau-1} \right], \\
\sum_{\tau\in\Theta_0} x_{t-\tau} &= \sum_{\tau\in\Theta_1\setminus\{1\}} x_{t-\tau+1} + \sum_{\tau\in\Theta_1} x_{t-\tau-1}.
\end{align}
\end{lemma}

\begin{lemma} \label{lem:singlesession}
Consider node $(0,u_2^d)\in\bar V$. Assume that the only non-zero source symbol transmitted in the network is $m^1_0(u_2^d)$ by node $(0,u_2^d)$ in time slot $0$. Then
\begin{equation}
x_t(\vv) = 
\begin{cases}
	m^1_0(u_2^d),\quad &\text{if } v_1=t, v_2^d=u_2^d, \\
	0,\quad &\text{otherwise,}
\end{cases}
\end{equation}
for all $v\in V$ and $t\geq 0$.
\end{lemma}
\begin{IEEEproof}
We use induction over $t$. At time $t=0$ the lemma holds, giving us our base case. Now suppose that the lemma holds for all time slots smaller than $t$. If $v\in\bar V$ the lemma follows directly from~(\ref{eq:rectcoding1})--(\ref{eq:rectcoding3}). In the remainder we consider $\uu\in\VVint$. From the induction hypothesis it follows that for any $t'<t$
\begin{equation} \label{eq:frominductionhyp}
	x_{t'}(\uu) = x_{t'-1}(u_1-1,u_2^d).
\end{equation}
If $u_1=K-1$, it follows from~(\ref{eq:rectcoding1}) and the induction hypothesis that
\begin{equation} \label{eq:frominductionhyp2}
	x_{t'-1}(\uu) = x_{t'}(u_1+1,u_2^d).
\end{equation}

Now, at $t$ the coding operation performed by $\uu$ can be decomposed as
\begin{align*}
x_t(\uu)
= &\ \sum_{\ww\in\NN_\uu\cup\{\vv\}}\sum_{\tau\in\Theta_{\dist(\ww,\uu)}}
	x_{t-\tau}(\ww)
= \sum_{\substack{\ww\in\NN_\uu:\\w_1=u_1}} g(\ww),
\end{align*}
where
\begin{multline}
g(\ww) = \sum_{\mathclap{\tau\in\Theta_{\dist(\ww,\uu)+1}}} x_{t-\tau}(w_1-1,w_2^d)
	+ \\ \sum_{\mathclap{\tau\in\Theta_{\dist(\ww,\uu)}}}x_{t-\tau}(\ww) + \sum_{\mathclap{\tau\in\Theta_{\dist(\ww,\uu)+1}}}x_{t-\tau}(w_1+1,w_2^d).
\end{multline}
In the remainder we show that
\begin{equation}
g(\ww) = \begin{cases}
x_{t-1}(w_1-1,w_2^d),\quad &\text{if } \ww=\uu\\
0,\quad &\text{otherwise,}
\end{cases}
\end{equation}
which proves the lemma, since by the induction hypothesis $x_{t-1}(u_1-1,u_2^d)=m^1_0(u_2^d)$ if $u_1=t$ and zero otherwise.

For $\ww\neq\uu$ we have
\begin{align*}
g(\ww)
= &\ \sum_{\mathclap{\tau\in\Theta_{\dist(\ww,\uu)+1}}}x_{t-\tau}(w_1-1,w_2^d) + \sum_{\mathclap{\tau\in\Theta_{\dist(\ww,\uu)}}}x_{t-\tau}(\ww) + \notag \\
	&\ \sum_{\mathclap{\tau\in\Theta_{\dist(\ww,\uu)+1}}}x_{t-\tau}(w_1+1,w_2^d) \\
= &\ \sum_{\mathclap{\tau\in\Theta_{\dist(\ww,\uu)+1}}}x_{t-\tau}(w_1-1,w_2^d) + \sum_{\mathclap{\tau\in\Theta_{\dist(\ww,\uu)+1}}}x_{t-\tau+1}(\ww) + \notag \\
	&\ \sum_{\mathclap{\tau\in\Theta_{\dist(\ww,\uu)+1}}}x_{t-\tau-1}(\ww) + \sum_{\mathclap{\tau\in\Theta_{\dist(\ww,\uu)+1}}}x_{t-\tau}(w_1+1,w_2^d) \\
= &\ \sum_{\mathclap{\tau\in\Theta_{\dist(\ww,\uu)+1}}}x_{t-\tau}(w_1-1,w_2^d) + \sum_{\mathclap{\tau\in\Theta_{\dist(\ww,\uu)+1}}}x_{t-\tau}(w_1-1,w_2^d) + \notag \\
	&\ \sum_{\mathclap{\tau\in\Theta_{\dist(\ww,\uu)+1}}}x_{t-\tau}(w_1+1,w_2^d) + \sum_{\mathclap{\tau\in\Theta_{\dist(\ww,\uu)+1}}}x_{t-\tau}(w_1+1,w_2^d) \\
= 0,
\end{align*}
where the second equality follows from Lemma~\ref{lem:recttheta}, the third equality follows from~(\ref{eq:frominductionhyp})--(\ref{eq:frominductionhyp2}) and the last equality holds because we work over $\mathbb{F}_2$.

For $\ww=\uu$ we have
\begin{align*}
g(\uu)
= &\ \sum_{\mathclap{\tau\in\Theta_1}}x_{t-\tau}(u_1-1,u_2^d) + \sum_{\mathclap{\tau\in\Theta_0}}x_{t-\tau}(\uu) + \notag \\
	&\ \sum_{\mathclap{\tau\in\Theta_1}}x_{t-\tau}(u_1+1,u_2^d) \\
= &\ \sum_{\mathclap{\tau\in\Theta_1}}x_{t-\tau}(u_1-1,u_2^d) + \sum_{\mathclap{\tau\in\Theta_1\setminus{\{1\}}}}x_{t-\tau+1}(\uu) + \notag \\
	&\ \sum_{\mathclap{\tau\in\Theta_1}}x_{t-\tau-1}(\uu) + \sum_{\mathclap{\tau\in\Theta_1}}x_{t-\tau}(w_1+1,w_2^d) \\
= &\ \sum_{\mathclap{\tau\in\Theta_1}}x_{t-\tau}(u_1-1,u_2^d) + \sum_{\mathclap{\tau\in\Theta_1\setminus{\{1\}}}}x_{t-\tau}(u_1-1,u_2^d) + \notag \\
	&\ \sum_{\mathclap{\tau\in\Theta_1}}x_{t-\tau}(u_1+1,u_2^d) + \sum_{\mathclap{\tau\in\Theta_1}}x_{t-\tau}(u_1+1,u_2^d) \\
= &\ x_{t-1}(u_1-1,u_2^d).
\end{align*}
\end{IEEEproof}

\begin{lemma} \label{lem:rectform}
Let $\uu\in\VVint$
\begin{equation}
x_t(\uu) = \sum_{i=1}^d \left[ m^i_{t-u_i}(\uum{i}) + m^{d+i}_{t-K+u_i}(\uum{i}) \right].
\end{equation}
\end{lemma}	
\begin{IEEEproof}
By linearity, time-invariance and symmetry of~(\ref{eq:rectcodingint}) together with Lemma~\ref{lem:singlesession}.
\end{IEEEproof}

We are now ready to prove that the destinations can correctly decode source symbols. We present the decoding procedure for nodes on the right border of the network, \ie for nodes of type $(K,u_2^d)\in\bar V$. The decoding procedures at the other borders can be obtained by exploiting the symmetry of the system.
\begin{lemma}
Consider node $\uu=(K,u_2^d)\in\bar V$. At the end of time slot $t-1$ it can decode symbol $m_{t-K}^1(u_2^d)$ as
\begin{multline} \label{eq:rectdec}
\sum_{\substack{\vv\in\NN_\uu:\\ v_1<K}}\sum_{\tau\in\Theta_{\dist(\uu,\vv)}} x_{t-\tau}(\vv) +
	\sum_{\substack{\vv\in\NN_\uu:\\ v_1=K}}\sum_{\tau\in\Theta_{\dist(\uu,\vv)+1}}[x^1_{t-\tau+1}(\vv) +\\ x^{d+1}_{t-\tau-1}(\vv)] +
	\sum_{\tau\in\Theta_{1}\setminus{\{1\}}}x^1_{t-\tau+1}(\uu) + \sum_{\tau\in\Theta_1}x^{d+1}_{t-\tau-1}(\uu)].
\end{multline}
\end{lemma}
\begin{IEEEproof}
First note that all terms in~(\ref{eq:rectdec}) correspond to symbols that have been received by $(K,u_2^d)$ before or in time slot $t-1$.

Now, from Lemma~\ref{lem:rectform} we have
\begin{align}
	\sum_{\substack{\vv\in\NN_\uu:\\ v_1<K}} & \sum_{\tau\in\Theta_{\dist(\uu,\vv)}} x_{t-\tau}(\vv) \notag \\
= &\ \sum_{\substack{\vv\in\NN_\uu:\\ v_1<K}}\sum_{\tau\in\Theta_{\dist(\uu,\vv)}}\sum_{i=1}^d \big[m_{t-v_i-\tau}^i(\vvm{i}) + m_{t-K+v_i-\tau}^{d+i}(\vvm{i}) \big] \notag \\
= &\ \sum_{\substack{\vv\in\NN_\uu:\\ v_1<K}}\sum_{\tau\in\Theta_{\dist(\uu,\vv)}} \big[m_{t-v_1-\tau}^1(\vvm{1}) + m_{t-K+v_1-\tau}^{d+1}(\vvm{1}) \big] + \notag \\
	&\ \sum_{i=2}^d\bigg[ \sum_{\substack{\vv\in\NN_\uu:\\ v_1<K, v_i=u_i}} \big[ \sum_{\tau\in\Theta_{\dist(\uu,\vv)+1}} m^i_{t-v_i+1-\tau}(\vvm{i}) + \notag \\
	&\ \sum_{\mathclap{\tau\in\Theta_{\dist(\uu,\vv)}}}m^i_{t-v_i-\tau}(\vvm{i}) + \sum_{\mathclap{\tau\in\Theta_{\dist(\uu,\vv)+1}}}m^i_{t-v_i-1-\tau}(\vvm{i}) + \notag \\
	&\ \sum_{\mathclap{\tau\in\Theta_{\dist(\uu,\vv)+1}}}m^{d+i}_{t-v_i+1-\tau}(\vvm{i}) + \sum_{\mathclap{\tau\in\Theta_{\dist(\uu,\vv)}}}m^{d+i}_{t-v_i-\tau}(\vvm{i}) + \notag \\ &\ \sum_{\mathclap{\tau\in\Theta_{\dist(\uu,\vv)+1}}}m^{d+i}_{t-v_i-1-\tau}(\vvm{i}) \big] \bigg] \notag \\
\overset{(a)}{=} &\ \sum_{\substack{\vv\in\NN_\uu:\\ v_1<K}}\sum_{\tau\in\Theta_{\dist(\uu,\vv)}} \big[m_{t-v_1-\tau}^1(\vvm{1}) + m_{t-K+v_1-\tau}^{d+1}(\vvm{1}) \big] \notag \\
= &\ \sum_{\tau\in\Theta_1} \big[m_{t-K+1-\tau}^1(\uum{1}) + m_{t-1-\tau}^{d+1}(\uum{1}) \big] + \notag \\
	&\ \sum_{\substack{\vv\in\NN_\uu:\\ v_1=K}}\sum_{\tau\in\Theta_{\dist(\uu,\vv)+1}} \big[m_{t-K+1-\tau}^1(\vvm{1}) + m_{t-1-\tau}^{d+1}(\vvm{1}) \big] \label{eq:rectdecodeproof1}
\end{align}
where $(a)$ holds, because for $\dist(\uu,\vv)>0$ Lemma~\ref{lem:recttheta} gives
\begin{multline*}
\sum_{\tau\in\Theta_{\dist(\uu,\vv)+1}} m^i_{t-v_i+1-\tau}(\vvm{i}) + \\
 \sum_{\mathclap{\tau\in\Theta_{\dist(\uu,\vv)}}}m^i_{t-v_i-\tau}(\vvm{i}) + \sum_{\mathclap{\tau\in\Theta_{\dist(\uu,\vv)+1}}}m^i_{t-v_i-1-\tau}(\vvm{i}) = 0
\end{multline*}
and
\begin{multline*}
\sum_{\mathclap{\tau\in\Theta_{\dist(\uu,\vv)+1}}}m^{d+i}_{t-v_i+1-\tau}(\vvm{i}) + \sum_{\mathclap{\tau\in\Theta_{\dist(\uu,\vv)}}}m^{d+i}_{t-v_i-\tau}(\vvm{i}) + \\ \sum_{\mathclap{\tau\in\Theta_{\dist(\uu,\vv)+1}}}m^{d+i}_{t-v_i-1-\tau}(\vvm{i}) = 0.
\end{multline*}

From~(\ref{eq:rectcoding1}) and~(\ref{eq:rectcoding2}) it follows that
\begin{multline}
\sum_{\substack{\vv\in\NN_\uu:\\ v_1=K}}\sum_{\tau\in\Theta_{\dist(\uu,\vv)+1}}[x^1_{t-\tau+1}(\vv) + x^{d+1}_{t-\tau-1}(\vv)] = \\
\sum_{\substack{\vv\in\NN_\uu:\\ v_1=K}}\sum_{\tau\in\Theta_{\dist(\uu,\vv)+1}} \big[m_{t-K+1-\tau}^1(\vvm{1}) + m_{t-1-\tau}^{d+1}(\vvm{1}) \big] \label{eq:rectdecodeproof2}
\end{multline}
and
\begin{multline}
\sum_{\tau\in\Theta_{1}\setminus{\{1\}}}x^1_{t-\tau+1}(\uu) + \sum_{\tau\in\Theta_1}x^{d+1}_{t-\tau-1}(\uu) = \\
\sum_{\tau\in\Theta_{1}\setminus{\{1\}}} m_{t-K+1-\tau}^1(\uum{1}) + \sum_{\tau\in\Theta_1} m_{t-1-\tau}^{d+1}(\uum{1}). \label{eq:rectdecodeproof3}
\end{multline}
The proof of the lemma follows by adding the final expressions from~(\ref{eq:rectdecodeproof1}), (\ref{eq:rectdecodeproof2}) and~(\ref{eq:rectdecodeproof3}) and observing that the outcome is $m_{t-K}^1(u_2^d)$.
\end{IEEEproof}

%
%
%
\subsection{Energy Consumption}
The energy consumption of the network coding scheme presented above provides an upper bound to $\min_r\Enc(V,M,r)$. 
\begin{lemma} \label{lem:rectenc}
$\Enc(V,M,\sqrt{d})\leq 4cd^{1+\alpha/2}(K-1)^{d-1}+cd^{\alpha/2}(K-1)^d$.
\end{lemma}	
\begin{IEEEproof}
All transmissions are over distance $\sqrt{d}$ and cost  $cd^{\alpha/2}$.
The nodes in $\bar V$ are transmitting twice. On each of the $2d$ sides of the network there are $(K-1)^{d-1}$ nodes from $\bar V$, hence $|\bar V|=2d(K-1)^{d-1}$. This gives $2|\bar V|=4d(K-1)^{d-1}$ transmissions. In addition, there are $(K-1)^d$ nodes in the interior, that are all transmitting once.
\end{IEEEproof}

Next, we give the minimum energy required by a routing solution.
\begin{lemma} \label{lem:recter}
$\Er(V,M,\sqrt{d}) = 2cd^{1+\alpha/2} \left\lceil K/\left\lfloor\sqrt{d}\right\rfloor\right\rceil (K-1)^{d-1}$.
\end{lemma}
\begin{IEEEproof}
Since, the transmission range is equal to $\sqrt{d}$, a routing solution requires $\lceil K/\lfloor\sqrt{d}\rfloor\rceil$ transmissions per session. Moreover, there are $|\bar V|=2d(K-1)^{d-1}$ sessions.
\end{IEEEproof}

Using the above two lemmas we are able to prove Theorem~\ref{th:fix}.
\begin{IEEEproof}[Proof of Theorem~\ref{th:fix}]
Lemmas~\ref{lem:rectenc} and~\ref{lem:recter} give
\begin{align}
\Bfix(d)
&\geq \lim_{K\to\infty} \frac{\Er(V,M,\sqrt{d})}{\Enc(V,M,\sqrt{d})} \notag \\
&\geq \lim_{K\to\infty}\frac{ 2cd^{1+\alpha/2} \left\lceil K/\left\lfloor\sqrt{d}\right\rfloor\right\rceil (K-1)^{d-1}}{cd^{\alpha/2}[4d(K-1)^{d-1}+(K-1)^d]} \notag \\
&= \frac{2d}{\left\lfloor\sqrt{d}\right\rfloor}.
\end{align}
\end{IEEEproof}

%
%
%
\section{Discussion} \label{sec:concl}
We have given several constructions of energy-efficient network codes. These constructions serve to show that compared to plain routing, network coding has the potential of reducing energy consumption in wireless networks. Since we have provided only codes that are based on a centralized design, it remains to be shown in future work if and how this potential can be exploited using practical codes. Moreover, it would also be of interest to consider the energy-benefit in topologies in which the nodes are not positioned at a lattice, for instance, random networks.

In this work we have provided lower bounds on the energy benefit of network coding for wireless multiple unicast. Another open problem is to find upper bounds on the benefit.

%
%
%
\bibliographystyle{IEEEtran}
\bibliography{IEEEabrv,it_energy}

\begin{thebibliography}{10}
\providecommand{\url}[1]{#1}
\csname url@samestyle\endcsname
\providecommand{\newblock}{\relax}
\providecommand{\bibinfo}[2]{#2}
\providecommand{\BIBentrySTDinterwordspacing}{\spaceskip=0pt\relax}
\providecommand{\BIBentryALTinterwordstretchfactor}{4}
\providecommand{\BIBentryALTinterwordspacing}{\spaceskip=\fontdimen2\font plus
\BIBentryALTinterwordstretchfactor\fontdimen3\font minus
  \fontdimen4\font\relax}
\providecommand{\BIBforeignlanguage}[2]{{%
\expandafter\ifx\csname l@#1\endcsname\relax
\typeout{** WARNING: IEEEtran.bst: No hyphenation pattern has been}%
\typeout{** loaded for the language `#1'. Using the pattern for}%
\typeout{** the default language instead.}%
\else
\language=\csname l@#1\endcsname
\fi
#2}}
\providecommand{\BIBdecl}{\relax}
\BIBdecl

\bibitem{chen98qosoverview}
S.~Chen and K.~Nahrstedt, ``An overview of quality-of-service routing for the
  next generation high-speed networks: Problems and solutions,'' \emph{IEEE
  Network, Special Issue on Transmission and Distribution of Digital Video},
  pp. 64--79, Nov./Dec. 1998.

\bibitem{chang98energy}
J.~Chang and L.~Tassiulas, ``Energy conserving routing in wireless ad hoc
  networks,'' in \emph{Proceedings of the Fifth Annual ACM/IEEE International
  Conference on Mobile Computing and Network (MobiCom)}, Dallas, TX, Aug. 1998.

\bibitem{rodoplu99}
V.~Rodoplu and T.~H. Meng, ``Minimum energy mobile wireless networks,''
  \emph{{IEEE} Journal on Selected Areas in Communications}, vol.~17, no.~8,
  pp. 1333--1344, 1999.

\bibitem{ramanathan02topology}
R.~Ramanathan and R.~Rosales-Hain, ``Topology control of multi-hop wireless
  networks using transmit power adjustment,'' in \emph{Proceedings of IEEE
  INFOCOM}, Tel Aviv, Israel, Mar. 2000.

\bibitem{jones01survey}
C.~Jones, K.~Sivalingam, P.~Agarwal, and J.~Chen, ``A survey of energy
  efficient network protocols for wireless and mobile networks,''
  \emph{ACM/Kluwer Wireless Networks}, vol.~7, no.~4, pp. 343--358, 2001.

\bibitem{saraydar02efficient}
C.~U. Saraydar, N.~B. Mandayam, and D.~J. Goodman, ``Efficient power control
  via pricing in wireless data networks,'' \emph{{IEEE} Trans. Commun.},
  vol.~50, pp. 291--303, 2002.

\bibitem{goldsmith02challenges}
A.~Goldsmith and S.~Wicker, ``Design challenges for energy-constrained ad hoc
  wireless networks,'' \emph{{IEEE} Trans. Wireless Commun.}, vol.~9, no.~4,
  pp. 8--27, 2002.

\bibitem{ahlswede00}
R.~Ahlswede, N.~Cai, S.-Y.~R. Li, and R.~W. Yeung, ``Network information
  flow,'' \emph{{IEEE} Trans. Inf. Theory}, vol.~46, no.~4, pp. 1204--1216,
  2000.

\bibitem{li03linear}
S.-Y. Li, R.~Yeung, and N.~Cai, ``Linear network coding,'' \emph{{IEEE} Trans.
  Inf. Theory}, vol.~49, no.~2, pp. 371--381, 2003.

\bibitem{koetter03algebraic}
R.~Koetter and M.~M{\'e}dard, ``An algebraic approach to network coding,''
  \emph{{IEEE/ACM} Trans. Netw.}, vol.~11, no.~5, pp. 782--795, 2003.

\bibitem{ho06}
T.~Ho, M.~Medard, R.~Koetter, D.~Karger, M.~Effros, J.~Shi, and B.~Leong, ``A
  random linear network coding approach to multicast,'' \emph{{IEEE} Trans.
  Inf. Theory}, vol.~52, no.~10, pp. 4413--4430, 2006.

\bibitem{nctheory}
R.~W. Yeung and N.~Cai, ``Network coding theory,'' \emph{Foundations and
  Trends{\textregistered} in Communications and Information Theory}, vol.~2,
  no. 4 and 5, pp. 241--381, 2006.

\bibitem{ncmonol1}
C.~Fragouli and E.~Soljanin, ``Network coding fundamentals,'' \emph{Foundations
  and Trends{\textregistered} in Networking}, vol.~2, no.~1, pp. 1--133, 2007.

\bibitem{ncmonol2}
------, ``Network coding applications,'' \emph{Foundations and
  Trends{\textregistered} in Networking}, vol.~2, no.~2, pp. 135--269, 2007.

\bibitem{lun06mincost}
D.~S. Lun, N.~Ratnakar, M.~M{\'e}dard, R.~Koetter, D.~Karger, T.~Ho, E.~Ahmed,
  and F.~Zhao, ``Minimum-cost multicast over coded packet networks,''
  \emph{{IEEE} Trans. Inf. Theory}, vol.~52, no.~6, pp. 2608--2623, 2006.

\bibitem{winter87steiner}
P.~Winter, ``Steiner problem in networks: a survey,'' \emph{Networks}, vol.~17,
  no.~2, pp. 129--167, 1987.

\bibitem{goel08bound}
A.~Goel and S.~Khanna, ``On the network coding advantage for wireless multicast
  in euclidean space,'' in \emph{Proceedings of the 7th international
  conference on Information processing in sensor networks}, 2008, pp. 64--69.

\bibitem{fragouli08ebu}
C.~Fragouli, J.~Widmer, and J.-Y. Le~Boudec, ``Efficient broadcasting using
  network coding,'' \emph{{IEEE/ACM} Trans. Netw.}, vol.~16, no.~2, pp.
  450--463, 2008.

\bibitem{widmer05extreme}
J.~Widmer and J.-Y. Le~Boudec, ``{Network coding for efficient communication in
  extreme networks},'' in \emph{Applications, Technologies, Architectures, and
  Protocols for Computer Communication}.\hskip 1em plus 0.5em minus 0.4em\relax
  ACM Press New York, NY, USA, 2005, pp. 284--291.

\bibitem{wu05ciss}
Y.~Wu, P.~A. Chou, and S.-Y. Kung, ``Information exchange in wireless networks
  with network coding and physical-layer broadcast,'' in \emph{Proc. 39th
  Annual Conference on Information Sciences and Systems {(CISS)}}, 2005.

\bibitem{effros06tiling}
M.~Effros, T.~Ho, and S.~Kim, ``A tiling approach to network code design for
  wireless networks,'' in \emph{Information Theory Workshop, 2006. {ITW} '06
  Punta del Este. {IEEE}}, 2006, pp. 62--66.

\bibitem{kim07low}
S.~Kim, M.~Effros, and T.~Ho, ``On low-power multiple unicast network coding
  over a wireless triangular grid,'' in \emph{Forty-Fifth Annual Allerton
  Conference on Communication, Control and Computing}, 2007.

\bibitem{kramer06ecb}
G.~Kramer and S.~A. Savari, ``Edge-cut bounds on network coding rates,''
  \emph{Journal of Network and Systems Management}, vol.~14, no.~1, pp. 49--67,
  2006.

\bibitem{keshavarzhaddad08}
A.~Keshavarz-Haddad and R.~Riedi, ``{Bounds on the Benefit of Network Coding:
  Throughput and Energy Saving in Wireless Networks},'' in \emph{Proc. of IEEE
  INFOCOM}, 2008, pp. 376--384.

\bibitem{katti06xor}
S.~Katti, H.~Rahul, W.~Hu, D.~Katabi, M.~M{\'e}dard, and J.~Crowcroft, ``{XOR}s
  in the air: practical wireless network coding,'' in \emph{Proc.\ of ACM
  SIGCOMM}, 2006, pp. 243--254.

\bibitem{liu07scale}
J.~Liu, D.~Goeckel, and D.~Towsley, ``{Bounds on the gain of network coding and
  broadcasting in wireless networks},'' in \emph{Proc. of IEEE INFOCOM}, 2007,
  pp. 6--12.

\end{thebibliography}

\end{document}